# Lab Manual of Introductory Physics-I for Virtual Teaching

*Neel Haldolaarachchige
Department of Physical Science, Bergen Community College, Paramus, NJ 07652

Kalani Hettiarachchilage
Department of Physics, Seton Hall University, South Orange, NJ 07962

December 12, 2020

*Corresponding Author: haldo.physics@gmail.com

**ABSTRACT**

A set of virtual experiments were designed to use with introductory physics I (analytical and general) class, which covers kinematics, Newton laws, energy, momentum and rotational dynamics. Virtual experiments were based on video analysis and simulations. Only open educational resources (OER) were used for experiments. Virtual experiments were designed to simulate in person physical lab experiments. All the calculations and data analysis (analytical and graphical) were done with Microsoft excel. Formatted excel tables were given to students and step by step calculations with excel were done during the class. Specific emphasis was given to student learning outcomes such as understand, apply, analyze and evaluate. Student learning outcomes were studied with detailed lab reports per each experiment and end of the semester written exam (which based on experiments). Lab class was fully web enhanced and managed by using Learning management system (LMS). Every lab class was recorded and added to the LMS. Virtual labs were done by using live video conference technology and labs were tested with both synchronous and asynchronous type of remote teaching methods.

**CONTENTS**

# EXPERIMENT 1

## PROPAGATION OF ERROR AND DATA ANALYSIS WITH MICROSOFT EXCEL

**OBJECTIVE**

Experimental uncertainties of measurements are investigated. Uncertainty of physical quantity is investigated with repetitive measurements. Uncertainty of calculated physical parameters is investigated by using propagation of error analysis.

**THEORY AND PHYSICAL PRINCIPLES**

***Significant Figures***

Units allow for a standard of measurement comparison. The standard form to express numbers is through the *International System of Units* (abbreviated SI). SI base units can simplify large and small quantities through scientific notation.

In physics-I class only following base SI units are used. Length is measured in meters (m), time is measured in second (s) and mass is measured in kilogram (kg).

Measured numbers almost always involve some degree of uncertainty. The number of *significant values* derived from a measuring instrument depends on its *least count*, which is the smallest interval that could be read from the instrument. Usually mechanical (non-digital) type instruments*,* it is possible to estimate the fraction of the least count of the instrument which is the uncertainty of the instrument.

More *significant figures* in a measurement signify greater reliability. The only zeros significant are those that are in between numbers and trailing zeros that are unnecessary to hold the decimal point. When calculating significant figures, round as many significant figures in the final answer as there are the least number of significant figures in its components.

***Experimental Uncertainty***

The two broad types of experimental uncertainty are *random and systematic errors*.

*Random errors,* also known as indeterminate and statistical, result from unexpected fluctuations that occur in experimental environments, leading to different values in repeated measurements.

*Systematic errors* occur due to the observer or the instruments being used. The same errors are obtained when measurement is repeated, and the source of error is identifiable.

*Accuracy* refers to how close a measurement approaches the true, accepted value. *Precision* displays how closely spread measurements come out to be.

Error of any single measurements is the *uncertainty of the instrument*. Which is depending on the least count of the instrument. For mechanical (non-digital) instruments it is possible to estimate the fraction of the least count of the instrument which is the uncertainty of the instrument. However, for consistency we will use the uncertainty of the mechanical instruments as half of the least-count of the instruments during this lab. This is valid only when a physical quantity is measured by using manual (non-digital) type instruments. If measured in a digital instrument, then the uncertainty is the least count of the instrument.

*For manual (non-digital) instruments,*

Uncertainty of Instrument = ½ * Least count of instrument (1)

*For digital instrument,*

Uncertainty of instrument = Least count of the instrument = Place value of the last decimal place (2)

*Fractional error* of single measurement = $\frac{\text{Uncertainty of instrument}}{\text{measured value}}$ (3)

*Percent error* of single measurement = $\frac{\text{Uncertainty of instrument}}{\text{measured value}} \times 100$ (4)

*Accepted value,* written as A, is the most accurate and official measurement quantity found in textbooks. *Experimental quantity* E is what is derived from personal measurement.

Absolute difference is the range between E and A. With the values of E and A, *fractional and percent error* can be determined:

$$\text{Fractional error} = \frac{|E-A|}{A} \quad (5)$$

$$\text{Percent error} = \frac{|E-A|}{A} \times 100\% \quad (6)$$

When there is no accepted value, it is still possible to compare results of two measurements $E_1$ and $E_2$ for the *percent difference*:

$$\text{Percent difference} = \frac{|E_2-E_1|}{(E_2+E_1)/2} \times 100\% \quad (7)$$

The *mean,* or the $\bar{x}$, represents the average of all the numerical data. The sum of all values is divided by the total number, n, in that set:

$$\bar{x} = \frac{x_1+x_2+x_3+\ldots+x_N}{n} = \frac{\sum_{i=1}^{i=n} x_i}{n} \quad (8)$$

*Deviation from the mean,* represented by $d_i$, is calculating how much an individual data $x_i$ differs from the mean value $\bar{x}$:

$$d_i = x_i - \bar{x} \quad (9)$$

Taking the absolute value of $d_i$ and calculating for *mean deviation* out of n measurements is the absolute mean error of the measurement.

$$\bar{d} = \frac{|d_1|+|d_2|+|d_3|+\cdots+|d_n|}{n} = \frac{\sum_{i=1}^{n}|d_i|}{n} \quad (10)$$

In this lab class, whenever many measurements for the single physical quantity is done, mean absolute deviation will be considered as the error of such measurement.

When many data are taken for one physical quantity then there are other standard methods to analyze the error of such data sets as shown below.

*Standard deviation (s) and variance ($s^2$)* can also represent the spread of individual data from the mean of n measurements.

$$s = \sqrt{\frac{\sum_{i=1}^{i=n}(x_i - \underline{x})^2}{n-1}} \qquad (11)$$

$$s^2 = \frac{\sum_{i=1}^{i=n}(x_i - \underline{x})^2}{n-1} \qquad (12)$$

***Propagation of Error***

Uncertainty of calculated quantities such as circumference, area and volume depend on error of more than one physical quantity. Then the question is how to calculate the uncertainty of circumference, area, and volume? Answer to this question is depending on how the individual physical quantities are combined in the equation or arithmetic of the equation. Uncertainty of such physical quantities can be found by partial differentiating the equation.

I. Addition or subtraction of measured quantities

Consider a calculated physical quantity such as circumference of a table, which involves addition of two directly, measured physical quantities of length and width.

Perimeter = P = 2L + 2W (13)

Uncertainty of Perimeter = $\delta P = \sqrt{(2\,\delta L)^2 + (2\,\delta W)^2}$ (14)

In general, for the uncertainty of addition of subtraction can be written as follows:

Z = A + B – D (15)

$$\delta Z = \sqrt{(\delta A)^2 + (\delta B)^2 + (\delta D)^2} \qquad (16)$$

II. Physical quantity with pre-multiplication factor and/or power factor

When a physical quantity involves a multiplication factor:

X = a Y (17)

$$\delta X = |a|\,\delta Y \qquad (18)$$

When a physical quantity involves a power factor:

$$X = a\,Y^n \qquad (19)$$

$$\frac{\delta X}{|X|} = |n|\,\frac{\delta Y}{|Y|} \qquad (20)$$

III. Multiplication and division of measured quantities

Consider a calculated physical quantity such as area of a table, which involves multiplication of two directly, measured physical quantities of length and width.

Area = A = L * W (21)

Fractional uncertainty of Area = $\frac{\delta A}{A} = \sqrt{\left(\frac{\delta L}{L}\right)^2 + \left(\frac{\delta W}{W}\right)^2}$ (22)

$$\text{Volume} = V = L * W * T \quad (23)$$

Fractional uncertainty of Volume = $\frac{\delta V}{V} = \sqrt{\left(\frac{\delta L}{L}\right)^2 + \left(\frac{\delta W}{W}\right)^2 + \left(\frac{\delta T}{T}\right)^2}$ (24)

In general, for the fractional uncertainty of multiplication and/or division can be written as follows:

$$Z = \frac{X\,Y\,Q}{P} \quad (25)$$

$$\frac{\delta Z}{Z} = \sqrt{\left(\frac{\delta X}{X}\right)^2 + \left(\frac{\delta Y}{Y}\right)^2 + \left(\frac{\delta Q}{Q}\right)^2 + \left(\frac{\delta P}{P}\right)^2} \quad (26)$$

**APPARATUS AND PROCEDURE**

- Measure length, width, and thickness of a table at your home. You can select any table to measure such as the dining table, working table, etc.
- Use a meter ruler or tape measure to do the measurements. If you do not have either of them then use a standard ruler such as 15 cm (half feet or 6 inches) ruler or 30cm (one feet or 12 inches) ruler.
- If you do not have any of the rulers in your house, then use a standard printing (or office) paper. Which is generally 27.9cm (11 inches) long.
- Must include a picture of each instrument used for this experiment such as the table, the measuring instrument and the laptop/PC showing excel software.
- Calculations and data analysis must be done with Microsoft Excel. Free student copy of Microsoft office can be downloaded here: https://www.microsoft.com/en-us/education/products/office
- A very detail video tutorial with data collection from simulator and data analysis with excel can be found here: https://youtu.be/nAuU5K_BS8Q

**PRE LAB QUESTIONS**

1) Why is it important to take multiple reading to accurately measure a physical quantity?
2) Explain the difference between precision and measuring accuracy?
3) Accuracy of the experiment is given by which error calculation, percent error or percent difference?
4) Precision of the experiment is given by which error calculation, percent error or percent difference?
5) How will you determine how many significant figures are needed to report after calculating a physical parameter which consist of many different measured parameters?

**DATA TABLES AND CALCULATIONS**

*Data tables and calculations must be done in excel. All the excel functions should be added below each table in the lab report.*

I. *Significant figures and uncertainty*

- Perimeter, Area and volume of table-top by using meter ruler
    - What is the least count of the meter ruler =
    - What is the uncertainty of the meter ruler =

Table 1 Dimensional measurements of table-top

| Trial | Length L [ ] | $\|L - \bar{L}\|$ [ ] | Width W [ ] | $\|W - \bar{W}\|$ [ ] | Thickness T [ ] | $\|T - \bar{T}\|$ [ ] |
|---|---|---|---|---|---|---|
| 1 | | | | | | |
| 2 | | | | | | |
| 3 | | | | | | |
| 4 | | | | | | |
| 5 | | | | | | |
| 6 | | | | | | |
| 7 | | | | | | |
| 8 | | | | | | |
| Average | | | | | | |

Table 2 Dimensional measurements of table-top

| Physical Quantity | Write in standard format with given units | Write in standard format with SI units |
|---|---|---|
| Average Length | | |
| Average Width | | |
| Average Thickness | | |

Table 3 Calculation of perimeter, area, and volume and uncertainties

| Average Perimeter | Uncertainty of average perimeter |
|---|---|
| | |
| Average Area | Uncertainty of average area |
| | |
| Average Volume | Uncertainty of average volume |
| | |

Table 4 Calculated physical parameters

| Physical Quantity | Write in standard format without rounding | Write in standard format with rounding |
|---|---|---|
| Perimeter | | |
| Area | | |
| Volume | | |

# EXPERIMENT 2 MEASURING INSTRUMENTS AND DENSITY

**OBJECTIVE**

Object dimensions are measured by using laboratory balance, a ruler, a Vernier caliper, and a Micrometer screw gauge. Using measurements with these instruments, objects volume and density are calculated. Experimental densities will then be compared to accepted values of density for the purpose of determining the possible material compositions of the objects. Propagation of errors are analyzed.

**THEORY AND PHYSICAL PRINCIPLES**

***Measuring Instruments***

Length is generally measured by using meter ruler and the least count (or the smallest measurement) of the meter ruler is one millimeter (1.0 mm). Therefore, meter rulers are only useful to measure length segments larger than one millimeter. There are two specific instruments that are used to measure length segments smaller than one millimeter, namely Vernier caliper and Micrometer screw gauge.

***Vernier Caliper***

Figure 1 shows a picture of Vernier caliper. There are two scales in the caliper. One is the main scale with the least count of one millimeter and the other is the Vernier scale which is the moving scale and the value of the one segment of the Vernier scale is the least count of the instrument.

To find the least count of the instrument, first close the jaws completely which means the external jaws should be touching each other. Then it should show zeros of both main and Vernier scales have aligned each other and read the main scale value in millimeters which aligned with last segment line of the Vernier scale.

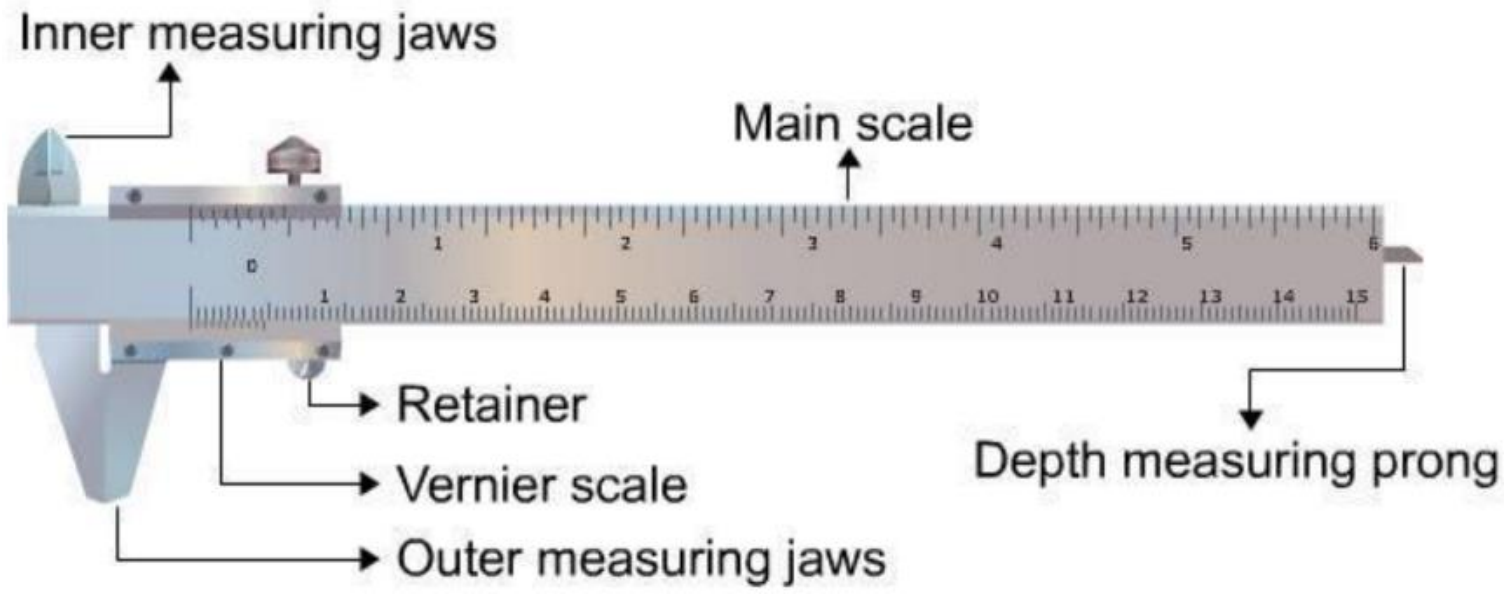

Figure 1 Vernier Caliper (Picture credit: https://amrita.olabs.edu)

$$\text{Least count of Vernier caliper} = \text{One main scale division} - \left(\frac{\text{main scale reading}}{\text{number of division in vernier scale}}\right) \quad (1)$$

$$\text{Least count of Vernier caliper} = \frac{\text{values of smalest segment of main scale}}{\text{number of division in vernier scale}} \quad (2)$$

If a Vernier caliper has 10 segments in the Vernier scale, the least count is 0.1 mm and if the Vernier scale consists of 20 segments then the least is 0.05 mm.

In general, instrumental uncertainty (error of a measurement) for analogue type instruments,

$$\text{Instrument uncertainty} = \frac{\text{least count}}{2} \quad (3)$$

Therefore, uncertainty of Vernier caliper with 0.1mm least count is 0.05 mm.

***Micrometer Screw Gauge***

Figure 2 shows a picture of a Micrometer screw gauge. There are two scales in the micrometer. One is the main scale with the least count of one millimeter (or some Micrometers with half a millimeter) and the other is circular scale which is the moving scale and the value of the one segment of the circular scale is the least count of the instrument.

To find the least count of the instrument, thimble should rotate one complete circle and find out how far the circular scale moves horizontally in terms of segments in the main scale. Usually when the thimble rotates one circle then the circular scale moves one segment (about 0.5 mm) of the main scale.

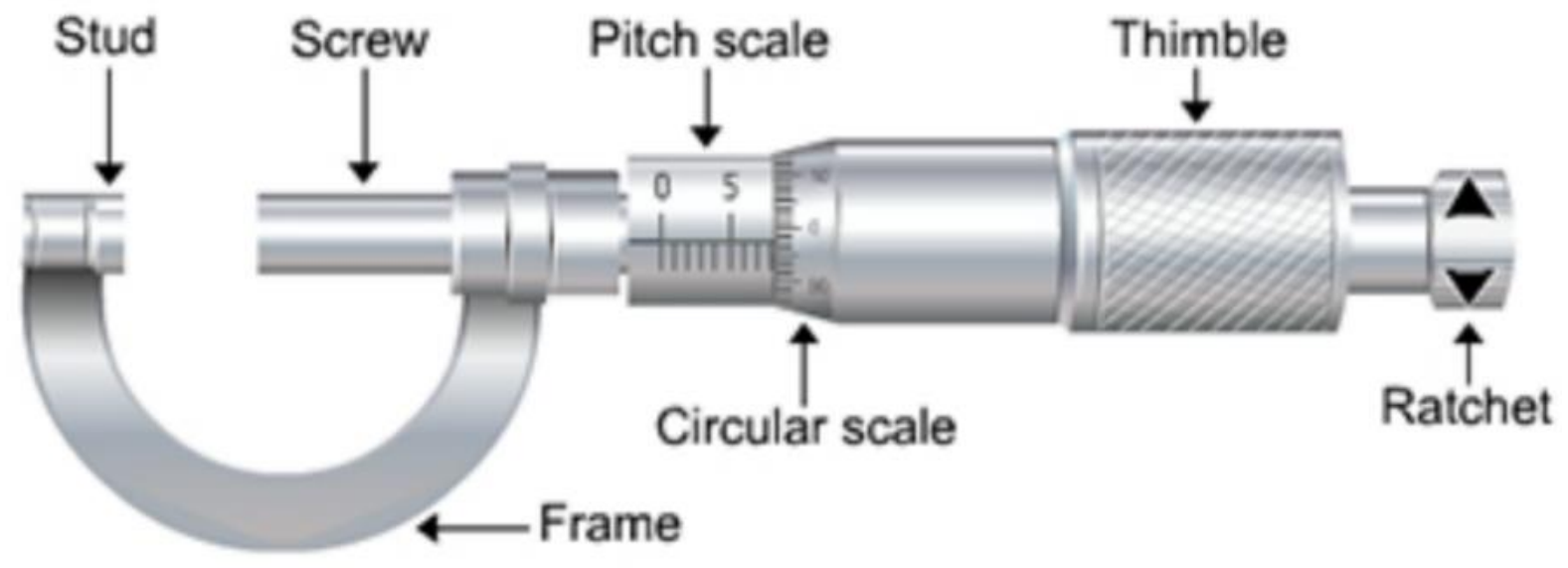

Figure 2 Micrometer Screw Gauge (Picture credit, https://amrita.olabs.edu)

$$\text{Least count of Micrometer} = \frac{\text{values of smalest segment of main scale}}{\text{number of division in circular scale}} \qquad (4)$$

If a circular scale has 100 segments, least count of the Micrometer is 0.01 mm. Therefore, the uncertainty of Micrometer is 0.005 mm.

In general, Micrometers are more accurate than the Vernier calipers. It is also important to note that data recording should indicate the uncertainty of the instruments. A measurement from the Vernier caliper of least count of 0.1mm should be recorded with two decimal places in which the second decimal place indicates the uncertainty of the instrument. Similarly, a measurement with a Micrometer of at least count 0.01mm should be recorded with three decimal places to properly indicate the instrument uncertainty.

Since Micrometer has circular scale which is a barrel moving on top of main scale it is very important to check the zero error. To check for zero error, first rotate the circular scale by using a ratchet till the screw touches the stud. When the screw and stud touch each other, the zero of the main scale should align with zero of circular scale. This is very important test that must be done before use of any Micrometer.

Figure 3 shows the two possible scenarios of zero errors. When two zeros are coincided then there is no zero error. If the circular scale zero is below the main scale horizontal line, then it is called negative zero error otherwise it is called positive zero error.

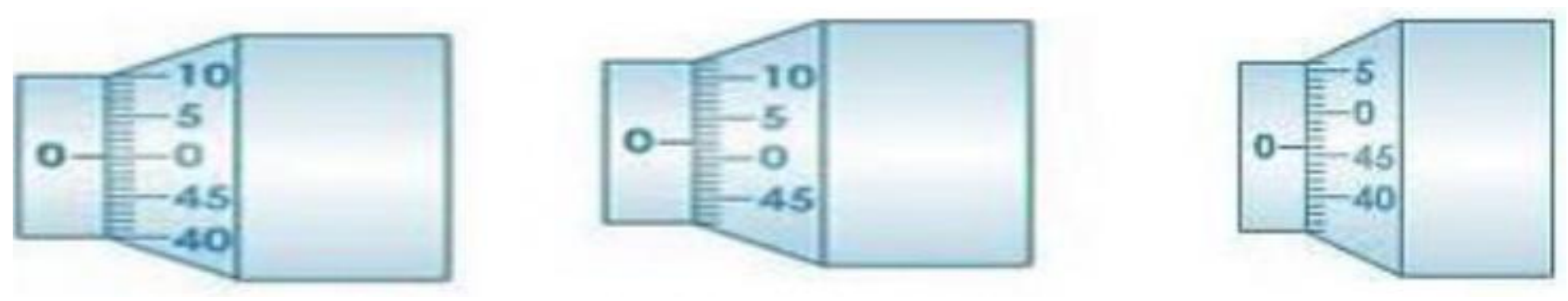

(a) (b) (c)

Figure 3 Zero-error of Micrometer Screw Gauge (Picture credit, https://amrita.olabs.edu)

Figure 3(a) two zeros are coincided therefore no zero error. In figure 3(b) circular scale zero is below the main scale zero and the number of segments from circular scale zero line to main scale horizontal line indicate the negative zero error. In figure 3(c) circular scale zero is above the main scale zero and the number of segments from circular scale zero line to main scale horizontal line indicate the positive zero error.

***Mass, Volume and Density***

The mass $M$ of an object is a measure of how much matter or how much material an object contains. Mass can be found using a laboratory balance which can be either mechanical or electronic. The volume $V$ of an object is a measure of the three-dimensional space of the object. The density $\rho$ of an object describes the compactness of matter inside the object or alternately describes how much matter is inside the three-dimensional space of the object. Density can be determined from the ratio of an object's mass to its volume $V$ and is given by

$$\rho = \frac{M}{V} \tag{5}$$

The SI units for density are $kg/m^3$ but $g/cm^3$ are commonly used for smaller values of density. Calculating the density of an object can help one to determine its material composition.

Volumes of a Plate, Cylinder, Sphere, and Irregularly Shaped Object

The volume of an object can be calculated in different ways, depending on the shape of the object.

The volume of a regularly shaped object can be determined from dimensional measurements. The volume of an irregularly shaped solid object can be determined by immersing it in a graduated cylinder with water. The displacement of the water level before and after the object is submerged reveals the volume of the object.

The volume $V$ of a rectangular plate can be found from measurements of length $L$, width $W$ and thickness $T$ and is given by

$$V = L \times W \times T \tag{6}$$

The volume $V$ of a solid cylinder or rod can be found by multiplying its circular cross-sectional area $A$ by its measured length $L$. The cross-sectional area $A$ depends on the measured diameter $D$ and is given by

$$A = \frac{\pi}{4} D^2 \tag{7}$$

Then the volume of a solid cylinder is

$$V = \frac{\pi}{4} D^2 L \tag{8}$$

The volume of a sphere can be determined by just a measurement of its diameter $D$ and is given by

$$V = \frac{\pi}{6} D^3 \tag{9}$$

Volume of an irregular shape object can be measured by using Archimedes principle. When an object is submerged into a water container, fluid volume displaced by the object is the same as the volume of the object.

Usually, fluid volume is measured in units of milliliters. Which must be converted into SI units of volume of cubic meters $m^3$.

$$1.0 \text{ mL} = 1.0 \text{ cm}^3 = 1.0*10^{-6} \text{ m}^3 = 1.0 \text{ L} = 10^{-3} \text{ m}^3 \quad (10)$$

***Propagation of Error and Uncertainty of Density Calculations***

Recall from *Experiment-01: Data and Error Analysis* that the uncertainty of a quantity $Q$ calculated with a mathematical form dependant on multiplication, division and/or powers of the measured quantities $X$, $Y$ and $Z$ such as

$$Q = \frac{X^a Y^b}{Z^c} \quad (11)$$

and where each measured quantity has its own individual uncertainty, is given by

$$\frac{\delta Q}{Q} = \sqrt{\left(a\frac{\Delta X}{X}\right)^2 + \left(b\frac{\Delta Y}{Y}\right)^2 + \left(c\frac{\Delta Z}{Z}\right)^2} \quad (12)$$

If multiple measurements of $X$, $Y$ and $Z$ were made, then $\Delta X$, $\Delta Y$ and $\Delta Z$ can be the mean absolute deviations or standard deviations of those respective measurements. Also $X$, $Y$, $Z$ and $Q$ in these equations can be replaced by avegrae values. The final calculated value for the calculated quantity $Q$ would then be reported as

$$Q = \bar{Q} \pm \delta Q \quad (13)$$

where $\bar{Q}$ is an average value of $Q$ calculated from the mean values$\bar{X}$, $\bar{Y}$ and $\bar{Z}$.

$$\bar{Q} = \frac{\bar{X}^a \bar{Y}^b}{\bar{Z}^c} \quad (14)$$

Fractional error of $\bar{Q}$

$$\frac{\delta Q}{\bar{Q}} = \sqrt{\left(a\frac{\Delta X}{\bar{X}}\right)^2 + \left(b\frac{\Delta Y}{\bar{Y}}\right)^2 + \left(c\frac{\Delta Z}{\bar{Z}}\right)^2} \quad (15)$$

fractional uncertainty of density for all shaped objects is given by

$$\frac{\delta \rho}{\bar{\rho}} = \sqrt{\left(\frac{\Delta M}{\bar{M}}\right)^2 + \left(\frac{\delta V}{\bar{V}}\right)^2} \quad (16)$$

Fractional uncertainty for the volume of a rectangular plate is thus,

$$\frac{\delta V}{\bar{V}} = \sqrt{\left(\frac{\delta L}{\bar{L}}\right)^2 + \left(\frac{\delta W}{\bar{W}}\right)^2 + \left(\frac{\delta T}{\bar{T}}\right)^2} \quad (17)$$

Fractional uncertainty of the volume of a solid cylinder is

$$\frac{\delta V}{\bar{V}} = \sqrt{\left(2\frac{\delta D}{\bar{D}}\right)^2 + \left(\frac{\delta L}{\bar{L}}\right)^2} \quad (18)$$

Fractional uncertainty of the volume of a sphere is

$$\frac{\delta V}{\bar{V}} = \sqrt{\left(3\frac{\delta D}{\bar{D}}\right)^2} = 3\frac{\delta D}{\bar{D}} \quad (19)$$

## APPARATUS AND PROCEDURE

### *Part A: Volume of an object by using Vernier caliper*

- This part of the experiment is done with following simulation: https://amrita.olabs.edu.in/?sub=1&brch=5&sim=16&cnt=4
- This simulation works fine with any web browser.
- A very detail video tutorial with data collection from simulator and data analysis with excel can be found here: https://youtu.be/i0F-28SU8K4

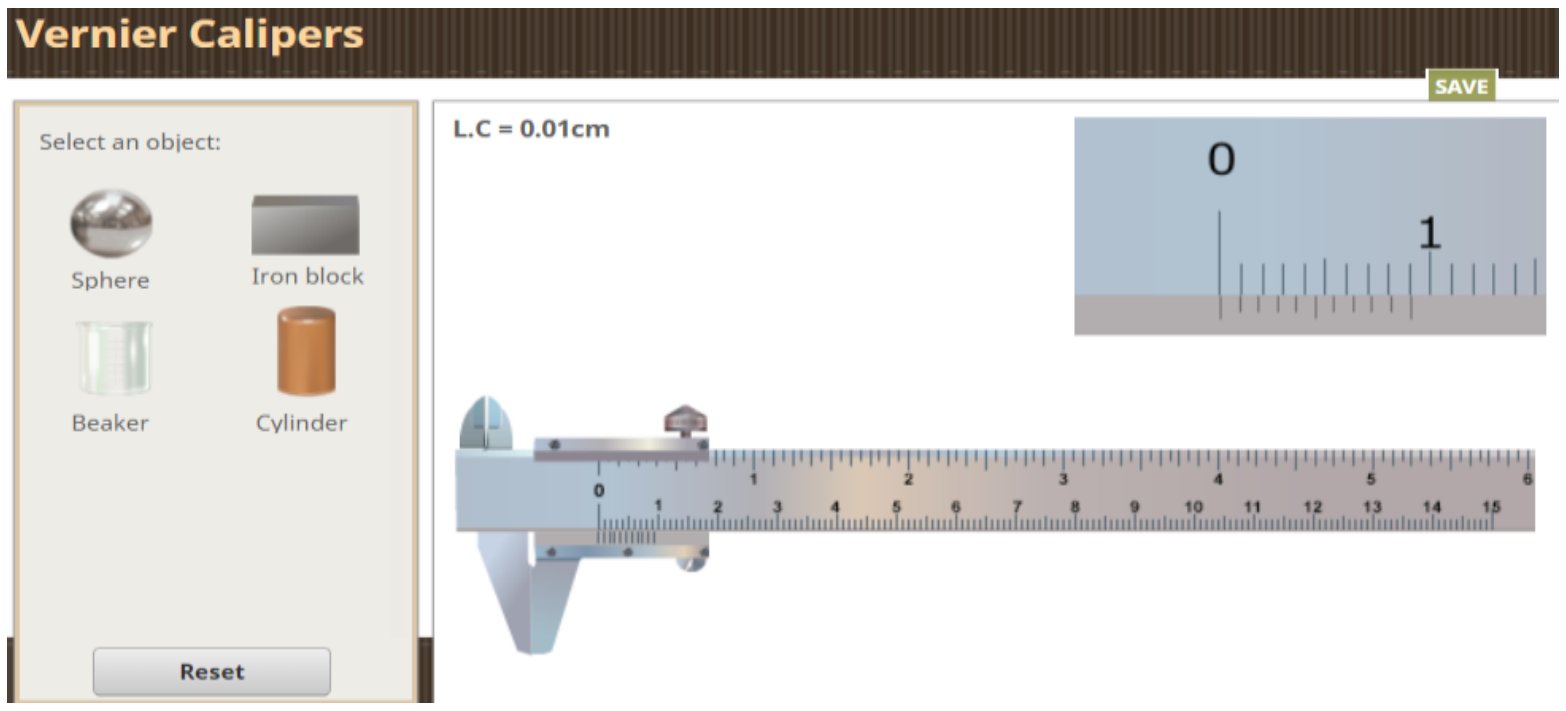

Figure 4 Vernier Caliper (Picture credit: https://amrita.olabs.edu)

- Click on the object to measure the dimension.
- Object appears on the vernier and the right outer jaw moves far right side.
- Click on the right outer jaw and drag it till it touches the object.
- When it is ready for reading then a zoomed in picture appears on top right side as shown in the figure.
- Read the measurement and record it in the data table.
- Repeat the above procedure as needed to measure dimensions of selected objects.

### *Part B: Volume of an object by using Micrometer Screw Gauge*

- This part of the experiment is done with following simulation: http://amrita.olabs.edu.in/?sub=1&brch=5&sim=156&cnt=4
- This simulation works fine with any web browser.
- A very detail video tutorial with data collection from simulator and data analysis with excel can be found here: https://youtu.be/i0F-28SU8K4

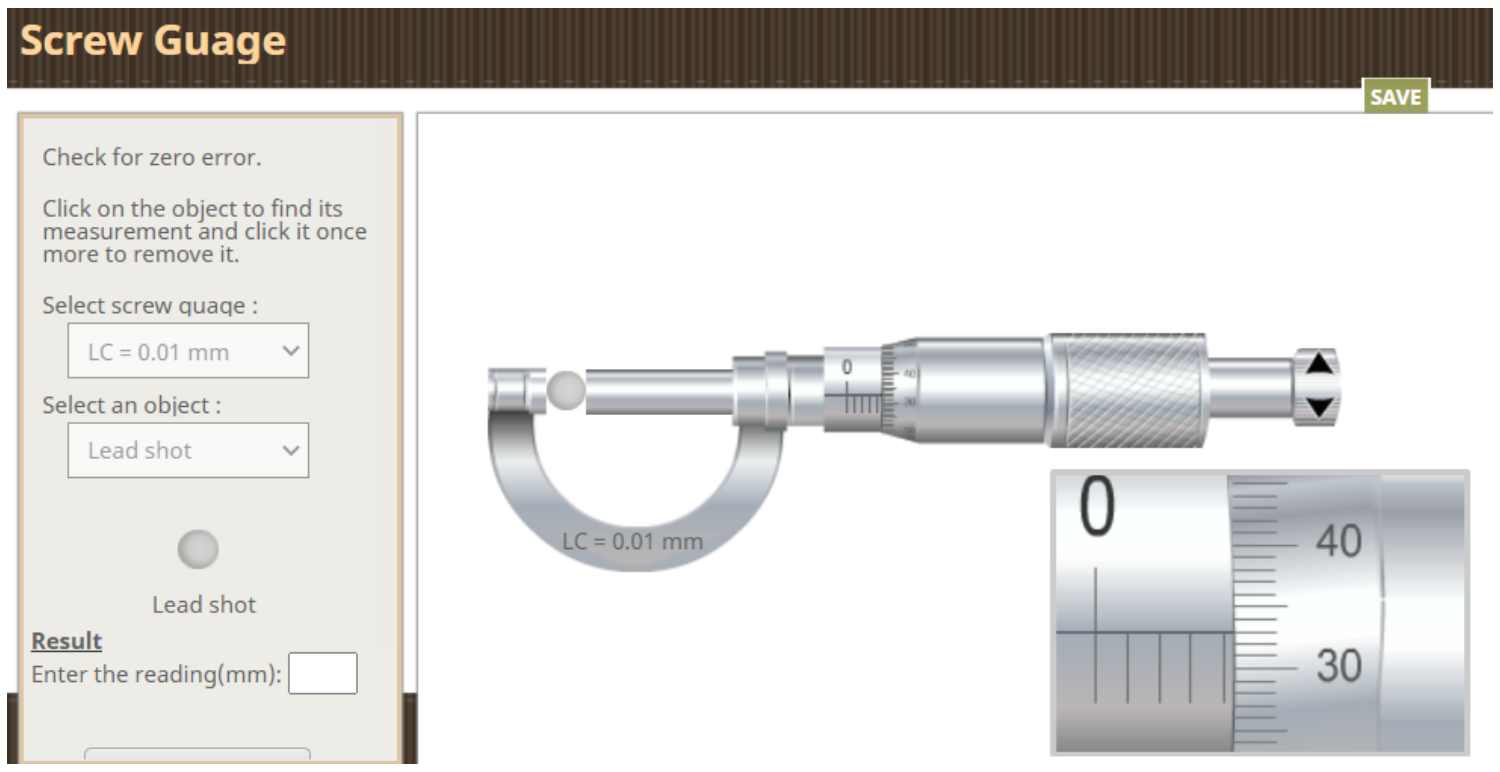

Figure 5 Micrometer Screw Gauge (Picture credit: http://amrita.olabs.edu.in)

- Zero error check must be done before use of micrometers.
- Click on the up arrow on the rachet end till the screw touches the stud.
- When screw and stud touch each other a zoomed in picture appears bottom right as shown in the figure.
- Check the zoomed in picture for possible zero error of the micrometer and record it.
- Click on the object to measure the dimension.
- Object appears on the micrometer and the screw moves far right side.
- Click on the black arrow on the rachet to move the screw till it touches the object.
- When it is ready for reading then a zoomed in picture appears on top right side as shown in the figure.
- Read the measurement and record it in the data table. Make sure to do any zero error correction to reading.
- Repeat the above procedure as needed to measure dimensions of selected objects.

***Part C: Density of an object***

- This part of the experiment is done with following simulation: http://amrita.olabs.edu.in/?sub=1&brch=1&sim=2&cnt=4
- This simulation works fine with any web browser.
- A very detail video tutorial with data collection from simulator and data analysis with excel can be found here: https://youtu.be/i0F-28SU8K4

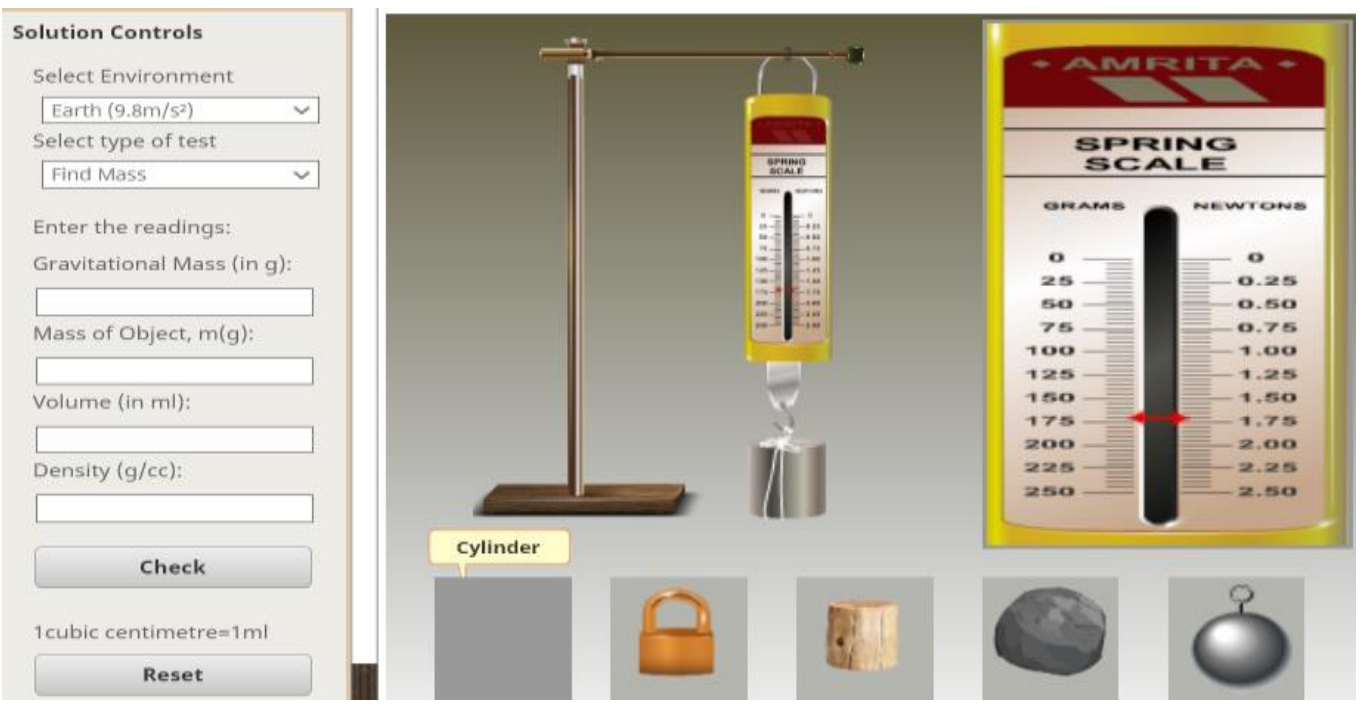

Figure 6 Micrometer Screw Gauge (Picture credit: http://amrita.olabs.edu.in)

- Mass of the object can be measured by using spring scale balance.
- It is important to check the least count of the balance, which can be done with the picture on the right side. There are two scales on the balance, Newton units on the right side and the kilogram units on the left side.
- Least count on the kilogram units of the balance is 5.0 grams. Therefore, uncertainty of the instrument is 2.5 grams.
- Click on the selected object on the bottom of the simulator.
- The object hangs on the scale and zoomed in picture on right can be used to read the mass of the object.
- To measure the volume of the object, reset the simulator by clicking the "reset" button on the bottom left side of the simulator.
- Select "find column" on the selection panel "select type of test" on the top left side of the simulator.
- A water cylinder appears. It is very important to check the least count of the water cylinder. Which is 5.0 mL. Therefore, the uncertainty of the instrument is 2.5 mL. Make sure to convert all the volume measurements to cubic centimeters or SI units of cubic meters.
- Repeat the measurement of mass and volume of any selected objects.

**PRE LAB QUESTIONS**

1) Describe two scales in Vernier Caliper?
2) Describe two scales in Micrometer screw gauge?
3) Describe a method to measure the thickness of a printing paper?
4) How do you measure the density of an irregular shape object?
5) Do you think the densities would change if measured at different temperatures?

Table 1 Standard Density Values of Selected Materials

| Name | Density [g/cm$^3$] | Name | Density [g/cm$^3$] |
|---|---|---|---|
| Aluminum | 2.70 | Calcite | 2.72 |
| Brass | 8.44 | Diamond | 3.52 |
| Copper | 8.95 | Feldspar | 2.57 |
| Iron | 7.86 | Halite | 2.12 |
| Lead | 11.48 | Magnetite | 5.18 |
| Nickel | 8.80 | Olivine | 3.32 |
| Silver | 10.49 | Cement | 1.85 |
| Tin | 7.10 | Chalk | 1.90 |
| Zinc | 6.92 | Clay | 1.80 |

## DATA ANALYSIS AND CALCULATIONS

*Data tables and calculations must be done in excel. All the excel functions should be added below each table in the lab report.*

### *Least Count and Uncertainty of Instruments*

Uncertainty of instrument = ½ * least count of instrument

Table 1 Uncertainty of Instruments

| Instrument | Least Count in given units | Least Count in SI units | Uncertainty of instrument in SI units |
|---|---|---|---|
| Meter Stick | | | |
| Vernier Caliper | | | |
| Micrometer Screw-Gauge | | | |
| Spring Scale Balance | | | |
| Water Cylinder | | | |

Vernier Caliper readings = Main scale reading + (Vernier scale reading x least count), should be in mm.

Micrometer readings = Main scale reading + (Micrometer scale reading x least count), should be in mm.

### *Part A: Volume of an object by using Vernier caliper*

Table 2 Dimension measurement of objects by Vernier

| Iron Block | | | |
|---|---|---|---|
| Dimension | Main scale [ ] | Vernier Scale [ ] | Final Reading [ ] |
| Length | | | |
| Width | | | |
| Thickness | | | |
| Metal Cylinder | | | |
| Dimension | Main scale [ ] | Vernier Scale [ ] | Final Reading [ ] |
| Length | | | |
| Diameter | | | |

Table 3 Object volume and uncertainty by Vernier

| Object | Volume equation | Volume calculated<br>V<br>[ ] | Uncertainty of volume equation | Uncertainty of Volume<br>$\delta V$<br>[ ] |
|---|---|---|---|---|
| Iron Block | | | | |
| Cylinder | | | | |

| | Volume calculated without rounding and in SI units<br>$V \pm \delta V$<br>[ ] | Volume calculated with rounding and in scientific notation and SI units<br>$V \pm \delta V$<br>[ ] |
|---|---|---|
| Iron Block | | |
| Cylinder | | |

***Part B: Volume of an object by using Micrometer Screw Gauge***

Table 4 Dimension measurement of objects by Micrometer

| Lead Shot | | | |
|---|---|---|---|
| Dimension | Main scale<br>[ ] | Circular Scale<br>[ ] | Final Reading<br>[ ] |
| Diameter | | | |
| Thin Wire | | | |
| Dimension | Main scale<br>[ ] | Circular Scale<br>[ ] | Final Reading<br>[ ] |
| Length | | | |
| Diameter | | | |

Table 5 Object volume and uncertainty by Micrometer

| Object | Volume equation | Volume calculated V [ ] | Uncertainty of volume equation | Uncertainty of Volume $\delta V$ [ ] |
|---|---|---|---|---|
| Lead shot | | | | |
| Thin wire | | | | |

| | Volume calculated without rounding and in SI units $V \pm \delta V$ [ ] | Volume calculated with rounding and in scientific notation and SI units $V \pm \delta V$ [ ] |
|---|---|---|
| Lead shot | | |
| Thin wire | | |

***Part C: Density of an object***

Table 6 Mass and volume measurements

| | Mass M [ ] | Uncertainty of Mass $(\Delta M)$ [ ] | Volume V [ ] | Uncertainty of Volume $\delta V$ [ ] |
|---|---|---|---|---|
| Cylinder | | | | |
| Lock | | | | |
| Stone | | | | |

| | Density $\rho$ [ ] | Uncertainty of Density | | |
|---|---|---|---|---|
| | | $\frac{\Delta M}{M}$ | $\frac{\delta V}{V}$ | $d\rho$ [ ] |
| Cylinder | | | | |
| Lock | | | | |
| Stone | | | | |

| | Density with calculated uncertainty $\rho \pm d\rho$ (*in SI units*) | Predicted Material | Expected Value of Density in SI unit | Percent Error (PE) |
|---|---|---|---|---|
| Cylinder | | | | |
| Lock | | | | |
| Stone | | | | |

# EXPERIMENT 3 VECTOR ADDITION AND VECTOR RESOLUTION

**OBJECTIVE**

- Vector addition by graphical and numerical methods.
- Use of cosine and sine rules to get the magnitude and direction of a resultant vector.
- Use of the vector component to find the magnitude and direction of the resultant vector.

**THEORY AND PHYSICAL PRINCIPLES**

***Vector addition with graphical method***

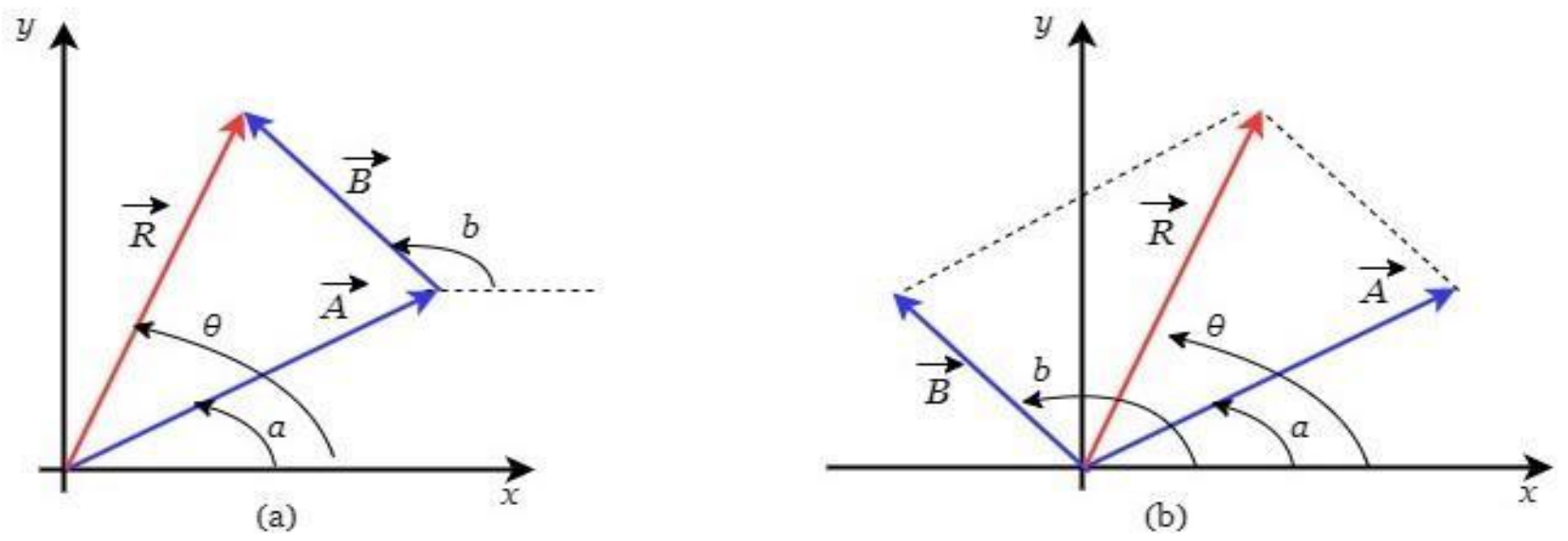

Figure 1 Vector-addition with triangular and parallelogram methods.

Vector addition can be done with graphically drawing the vectors in a graph sheet with the scale conversion from the given units to a scale in the graph sheet (mm or cm). There are two standard methods to find the resultant vector of addition of two vectors, namely a) triangular or tail-to-tip (head) and b) parallelogram as shown in the figure 1.

When two vectors $\vec{A}$ and $\vec{B}$ are added graphically, the vectors are drawn as arrows head-to-tail, in which a line from the tail of the first vector to the head of the second vector gives the resultant vector $\vec{R}$.

$$\vec{R} = \vec{A} + \vec{B} \tag{1}$$

The commutative law applies to vectors, in which the order added does not matter.

$$\vec{A} + \vec{B} = \vec{B} + \vec{A} \tag{2}$$

When given an additional vector $\vec{C}$, the associative law applies too.

$$(\vec{A} + \vec{B}) + \vec{C} = \vec{A} + (\vec{B} + \vec{C}) \tag{3}$$

The difference of vectors $\vec{A}$ and $\vec{B}$ can be found graphically by putting them into an additional form.

$$\vec{A} - \vec{B} = \vec{A} + (-\vec{B}) \tag{4}$$

***Vector addition with cosine and sine rules***

The resultant vector (R) from 2 vectors that do not form a right triangle can be analytically computed from the magnitude of the two known vectors ($F_1$ and $F_2$) and the angle between them (b) through the law of cosines.

$$R^2 = F_1^2 + F_2^2 - 2F_1F_2cosb \tag{5}$$

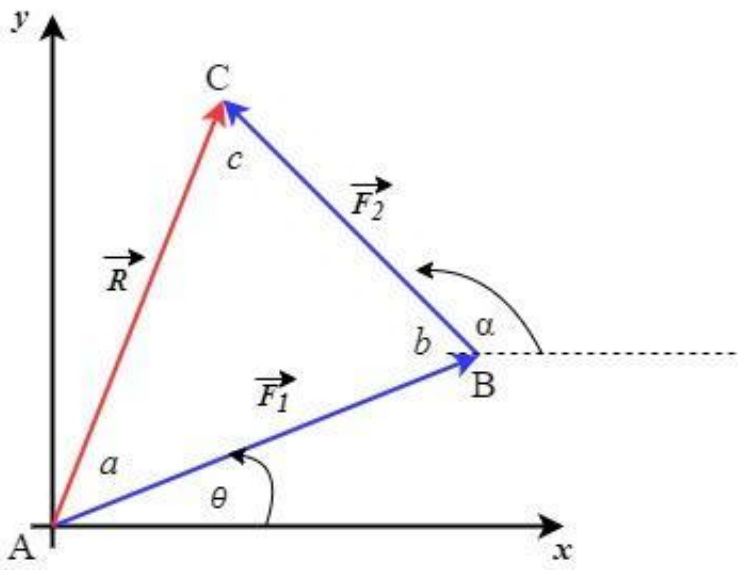

Figure 2 Vector-addition with triangular method

Consider any two forces F1 and F2, which are on xy plane and magnitude and directions are given as shown in the figure 1. Triangle ABC is the result of tail-to-tip graphical representation of vector addition. When making tri angular vector addition unit conversion from given unit of vectors to drawing units (cm) in the paper is required. Vector directions are relative to +x axis, which is the standard representation of vectors.

Angle b can be found by knowing angles of ɵ and α:

$$b = 180 - \alpha + \theta \quad (6)$$

Solving equation (5) for the resultant vector will be:

$$R = \sqrt{F_1^2 + F_2^2 - 2F_1F_2 cosb} \quad (7)$$

Direction of resultant vector R is, $\beta = \theta + a$ (8)

To find angle a, law of sines can be used,

$$\frac{sin\ c}{F_1} = \frac{sin\ a}{F_2} = \frac{sin\ b}{R} \quad (9)$$

$$b = sin^{-1}\left(\frac{R\ sin\ a}{F_2}\right) \quad (10)$$

***Vector addition with victor-component (projections on x and y axis)***

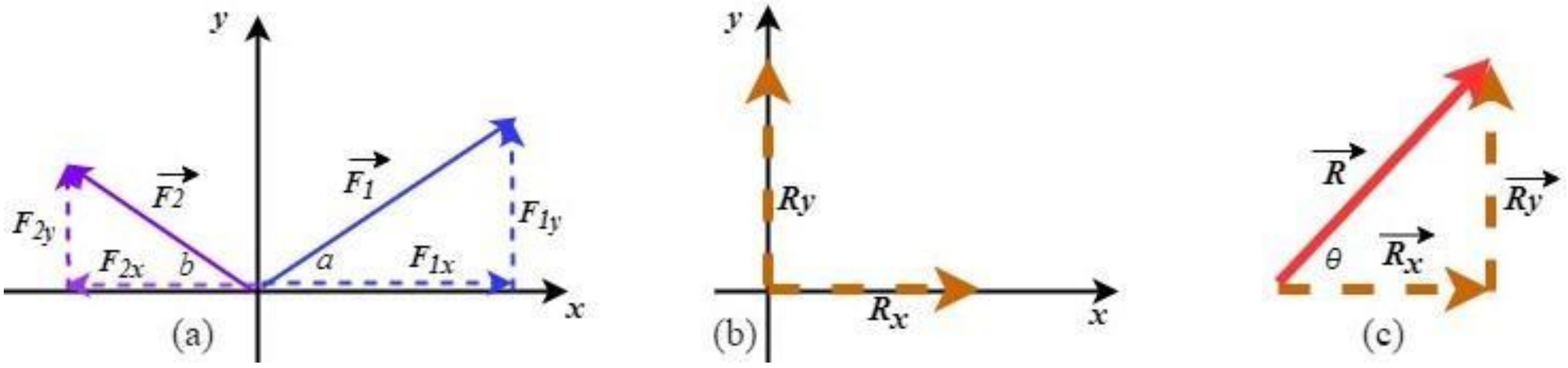

Figure 3 Vector-addition with vector-component method

With the component method, each vector (F) is separated into 2 components (projections) on the x ($F_x$) and y axis ($F_y$). The x-component ($F_{1x}$) of a vector-1 can be computed given the magnitude of the vector ($F_1$) and the angle it creates with the horizontal x-axis (a).

$$F_{1x} = F_1\ cos\ a \quad (11)$$

The y-component ($F_{1y}$) of a vector-1 can also be computed given the magnitude of the vector ($F_1$) and the angle it creates with the horizontal x-axis (a).

$$F_{1y} = F_1\ sin\ a \quad (12)$$

When adding multiple vectors, x-component of the resultant vector ($R_x$) is found by the x-components of vector 1 and vector 2.

$$\vec{R}_x = (F_{1x} - F_{2x})\hat{\imath} \tag{13}$$

The y-component of the resultant vector ($R_y$) is found by the y-components of vector 1 ($F_{1y}$) and vector 2 ($F_{2y}$).

$$\vec{R}_y = \left(F_{1y} + F_{2y}\right)\hat{\jmath} \tag{14}$$

The magnitude of the resultant vector (R), given the resultant x ($R_x$ ) and y-components ($R_y$), is found by:

$$R = \sqrt{R_x^2 + R_y^2} \tag{15}$$

The angle of the resultant vector from the horizontal x-axis (θ) can also be given through the resultant x-component ($R_x$) and y-components ($R_y$),

$$\theta = tan^{-1}\left(\frac{R_y}{R_x}\right) \tag{16}$$

The precision between measured values can be calculated by percent difference, using experimental values $E_1$ and $E_2$:

$$\text{Percent difference} = \frac{|E_2 - E_1|}{(E_2 + E_1)/2} \times 100\% \tag{17}$$

**APPARATUS AND PROCEDURE**

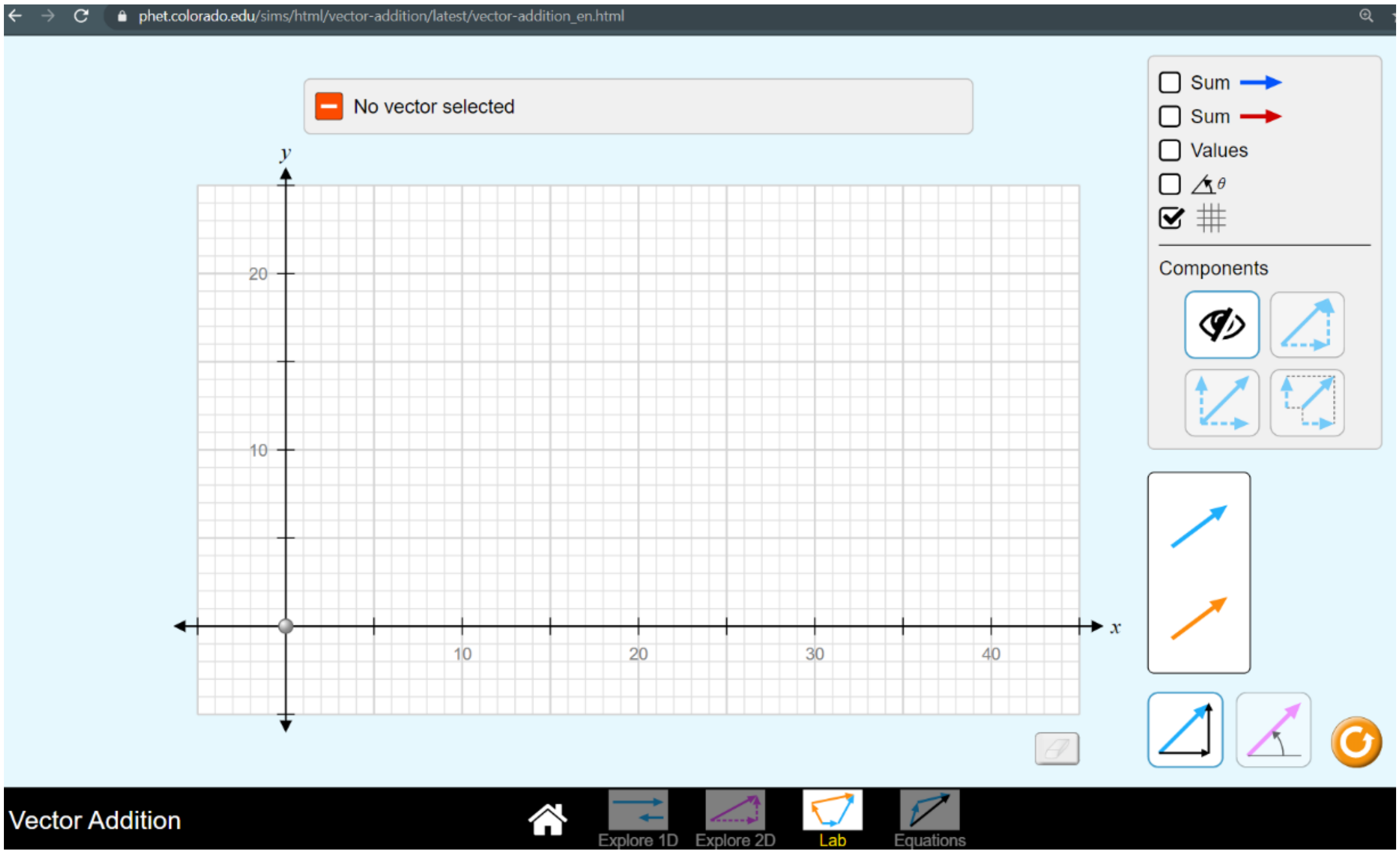

Figure 4 Vector addition simulation with PhET simulator (Picture credit, https://phet.colorado.edu)

- This simulation will run on a web browser (Chrome or Firefox is better) and if you prefer then there is an option to download the java-applet. Here is the direct link to projectile simulation. https://phet.colorado.edu/sims/html/vector-addition/latest/vector-addition_en.html
- A very detail video tutorial with data collection from simulator and data analysis with excel can be found here: https://youtu.be/uxiap4y4owU
- Vector addition with graphical methods can be one with the simulation.
- Click the simulation link above and select the "lab" to work on a graph sheet.
- Each vector can be drawn on the graph sheet by using the vector symbol in the right bottom of the simulation page (blue or orange arrows).
- Tail-to-tip (head) or triangular method is done with vectors drawn one after the other. Second vector should be started at the end of the first vector and the angle is measured from +x direction.
- Given vectors in the table-1 are in units of Newton and it can be converted into drawing units of 1.0mm=1.0N.
- Resultant vectors can be found by using the selection on the right-top of the simulation page or use a new vector to connect the origin (tail of the first vector) and end (tip or head) of the last vector.
- Direction of the resultant vector also can be found by using "angle" on the right-top of the simulation page.
- The other graphical vector addition is the parallelogram method. In this method all the vectors are drawn starting from the origin. This method should be done with only two vectors at a time.
- After two vectors are drawn from the origin, the parallelogram should be completed. Diagonal of the parallelogram is the resultant vector.
- When doing the three-vector addition, it is easy if you use the resultant vector of the first two and the third vector. Then it is like adding two vectors.
- When adding 4 vectors then can consider the resultant of the first three vectors and the last vector.
- Analytical method-I is done with cosine/sine rules. Write down all the detailed calculations in paper and pictures of your work should be attached with appendix C and final results should be summarized into table 3.
- Analytical method-II is done with component method and detail work should be included in appendix C and results should be summarized to table 3.

**PRE LAB QUESTIONS**

1) Describe the graphical vector addition with tail to head method?
2) Describe the graphical vector addition with parallelogram method?
3) Describe the sine and cosine rules?
4) Describe the vector addition with the vector component method?
5) Describe resultant and equilibrium vectors?

## DATA TABLES AND CALCULATIONS

*Data tables and calculations must be done in excel. All the excel functions should be added below each table in the lab report.*

- Vector addition with graphical methods can be done with the simulation.

Table 1 Results of vector addition by graphical methods

| | **Forces** | **Graphical Method I Tail-to-head (triangle)** | **Graphical Method II Parallelogram** |
|---|---|---|---|
| Vector addition A | $F_1 = 8.60N,\ \theta_1 = 35.5^0$<br>$F_2 = 11.7N, \theta_2 = 160^0$ | R =<br>$\theta =$ | R =<br>$\theta =$ |
| Vector addition B | $F_1 = 8.60N, \theta_1 = 35.5^0$<br>$F_2 = 11.7N, \theta_2 = 160^0$<br>$F_3 = 13.6N, \theta_3 = 234^0$ | R =<br>$\theta =$ | R =<br>$\theta =$ |
| Vector diagram of graphical method I of case A | | Vector diagram of graphical method II of case A | |
| | | | |
| Vector diagram of graphical method I of case B | | Vector diagram of graphical method II of case B | |
| | | | |

- Analytical methods should be done in a paper and scan them with a doc scanner in your phone and then attached them into report.
- Detail calculation work of the graphical and analytical methods should be included.
- Data tables are to summarize the final results.

Table 2 Results of vector addition by analytical methods

| | **Forces** | **Analytical Method I Cosine/sine rules** | **Analytical Method II Component** |
|---|---|---|---|
| Vector addition A | $F_1 = 8.60N,\ \theta_1 = 35.5^0$<br>$F_2 = 11.7N, \theta_2 = 160^0$ | R =<br>$\theta =$ | R =<br>$\theta =$ |
| Vector addition B | $F_1 = 8.60N, \theta_1 = 35.5^0$<br>$F_2 = 11.7N, \theta_2 = 160^0$<br>$F_3 = 13.6N, \theta_3 = 234^0$ | R =<br>$\theta =$ | R =<br>$\theta =$ |

Table 3 Comparison of results between different methods

| | **Forces** | **PD of Analytical I. and Analytical II** | **PD of Graphical I and Analytical I** | **PD of Graphical I and Analytical II** |
|---|---|---|---|---|
| Vector addi. A | $F_1 = 8.60N,\ \theta_1 = 35.5^0$<br>$F_2 = 11.7N, \theta_2 = 160^0$ | PD of R's =<br>PD of $\theta's$ = | PD of R's =<br>PD of $\theta's$ = | PD of R's =<br>PD of $\theta's$ = |
| Vector addi. B | $F_1 = 8.60N, \theta_1 = 35.5^0$<br>$F_2 = 11.7N, \theta_2 = 160^0$<br>$F_3 = 13.6N, \theta_3 = 234^0$ | PD of R's =<br>PD of $\theta's$ = | PD of R's =<br>PD of $\theta's$ = | PD of R's =<br>PD of $\theta's$ = |

# EXPERIMENT 4 UNIFORMLY ACCELERATED MOTION

## OBJECTIVE

Understand the gravitational field and its effect on freely falling objects. Measure gravitational acceleration by using free-falling objects. Understand the effect of drag force on free falling objects. Measure terminal velocity of free-falling objects.

## THEORY AND PHYSICAL PRINCIPLES

The displacement (Δy) of an object is the directional change in its position between position 1 ($y_1$) and position 2 ($y_2$).

$$\Delta y = y_2 - y_1 \tag{1}$$

The changing velocity (Δv) between location-1 and 2: $\Delta v = v_2 - v_1$ (2)

The changing time (t) between location-1 and 2: $t = t_2 - t_1$ (3)

The displacement (y) an object falls during a given time period (t) is given by the initial position ($y_0$), initial velocity ($v_0$), gravitational acceleration ($g$), and the time elapsed (t).

$$y = y_0 + v_0 t + \frac{1}{2} g t^2 \tag{4}$$

When object is dropped from height y above the ground,

$$y = \frac{1}{2} g t^2 \tag{5}$$

Final velocity (v) of an object falling at constant acceleration is given by its initial velocity ($v_0$), rate of gravitational acceleration ($g$), and time elapsed (t).

$$v = v_0 + gt \tag{6}$$

If the object begins from rest, $v = gt$ (7)

Gravitational constant can be found, $g = \frac{\Delta v}{\Delta t}$ (8)

The average velocity $\bar{v}$ of a uniformly accelerated object in a free fall is given by the instantaneous velocities at times $t_i$ ($v_i$) and $t_0$ ($v_0$) divided by 2.

$$\bar{v} = \frac{v + v_0}{2} = \frac{y}{t} \tag{10}$$

$$v = \frac{2y}{t} - v_0 \tag{11}$$

If the object falls from rest (dropped), $v = \frac{2y}{t}$ (12)

Average acceleration ($g$) is given by:

$$g = \frac{\Delta v}{\Delta t} = \frac{v_2 - v_1}{t_2 - t_1} \tag{13}$$

The accuracy of a measured value (E) compared to an accepted value (A) is given by the percent error (PE).

$$PE = \frac{|A-E|}{A} \times 100\% \quad (14)$$

*Terminal velocity and free fall*

When an object is in free fall that object experiences an upward force called drag force due to air flow. Drag force is related to the object velocity.

$$F_{drag} \propto v^2 \quad (15)$$

Velocity of free-falling objects increase due to gravitational acceleration and drag force also increase. If drag force becomes equal to the weight of the object then the object does not experience net force as a result the object does not experience an acceleration, so the object will experience weightlessness. After that point, the object will fall with constant velocity which is called terminal velocity.

## APPARATUS AND PROCEDURE

- Virtual experiment is done with tracker video analysis software. Which can be downloaded here: https://physlets.org/tracker/
- Tracker software should download and install into your laptop/PC. This software is compatible with all the operating systems windows/mac/linux.
- Here is the video tutorial that shows details of data collection with tracker and data analysis (analytical and graphing) with excel: https://youtu.be/V_qrdMwUF8M
- Drop three objects (golf ball, tennis ball and coffee filter) about 1.829m (6.0 feet) height from ground level) and record the video with slow-motion mode on a mobile phone.

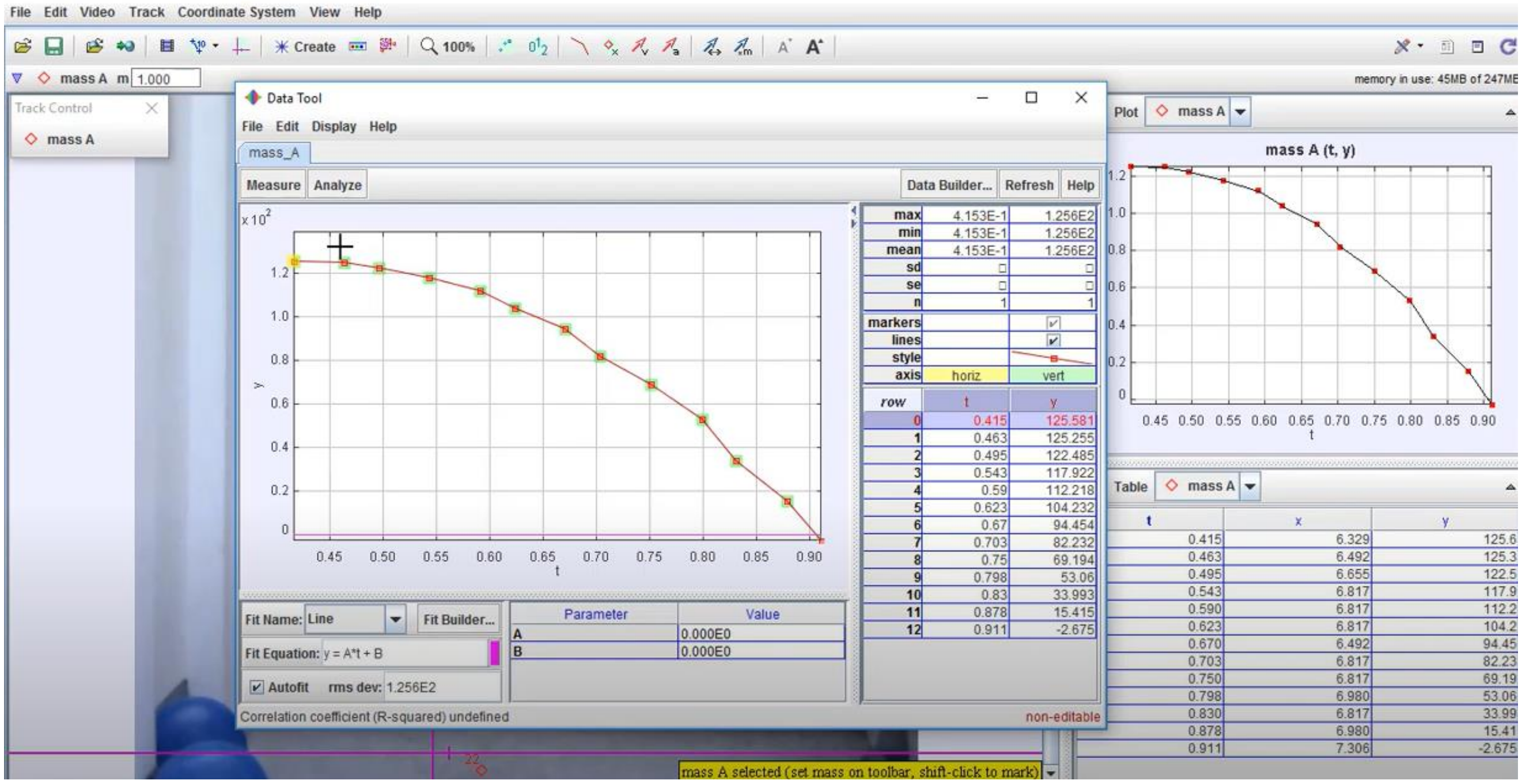

Figure 1 Tracker, a video analysis software for free-fall lab (Picture credit, https://physlets.org/tracker/)

***Video analysis and data collection of an object drop from 6.0 feet above ground***

- Import the golf ball video to the tracker. To import, click “video” on the menu bar and select the video.
- If the frame duration message shows up, then click “ok”.
- Right click on the video → click on “filters” → click “new” → click “rotate” → click “-90” degrees rotation.
- Right click again on the video → click on “clip setting” → change frame rate to 240 /s
- Click on “coordinate axis” on the menu bar. Which is the pick color coordinate axis symbol on the menu bar. Then pick color x and y axis appear on the video.
- Click on the origin point and drag it till the x axis is aligned with the golf ball.
- Click on white color arrow on the right to the green color play button at the bottom of the tracker then drag it till the frame number (red color number at the very left bottom) reaches close to 354.
- Make sure the x axis passes through the center of the golf ball.
- Click on “calibration tool” on the top menu bar. Which is the Blue color icon just left to the coordinate axis tool.
- Calibration tool → new calibration stick
- Hit “shift” and hold it then mouse icon changes into blue color square shape → bring the mouse on top of origin point then “click”. To work properly it should be shift + click tighter.
- Then you will see a number inside the blue square in the middle of the blue vertical line. Click on that square (one click) then change the number into 1.829 m.
- Click on “create” → click on “point mass” on top menu bar.
- Then you will see a small graph appear on the top right corner. And a small box with mass A term on top left.
- Click on the x(m) term vertical side of the graph on top right side. Then the selection panel appears and clicks on “y position, y component”.
- Check the frame number on bottom left and make sure that number is close to 354. If not, adjust the arrow on the play bar at the bottom of the video. Then click on the frame number and make it to zero.
- Hit “shift” then the mouse icon changes into blue color square shape. Then bring it on top of the ball and click. To work this properly it should be shift + click together.
- Then you will see a red color square appear on top of the ball. Click on the red color square. Few new icons (X, Y, r, theta) appear on just below the menu bar.
- Click on the white space on Y and delete all in it and type zero. This will make the starting y passion to y=0.
- Then click on the “forward” button till the ball moves a little bit from the previous position. Then do the shift + click on top of the ball.
- Then repeat the last two steps till the ball moves closer to ground. But collection of the data mist stops just before the ball hits the ground. Data should not be collected on or after the ball hits the ground.
- Right click on the graph on the top right corner. Click on the “analyze” on the selection panel.
- New window will have data and the graph. Then select all the data (Y and t columns) on the right side. Copy them and paste them into the excel file.

***Part A and B Data analysis of baseball and tennis ball drop from 6.0 feet above ground***

- Make a graph of displacement $y_i$ vs time $t_i$.
- Do the quadratic fitting and calculate the gravitational constant by using the factor in front of the quadratic term. Then, compare (find percentage error) with $g$ = -9.81 m/s$^2$.
- Then, make another graph of velocity $v_i$ vs time $t_i$.
- Do the linear fitting and find the gravitational constant by using the slope of the graph. Then, compare (find percentage error) it with $g$ = -9.81 m/s$^2$.

***Part C Data analysis of coffee filter drop from 6.0 feet above ground***

- Make a graph of displacement $y_i$ vs time $t_i$.
- First check the behavior of this data with quadratic fitting. If the object achieved terminal velocity then the data does not fit well quadratic fitting.
- Select the last few data points (closer to ground) and do the fitting with a linear trend line.
- Constant in front of the x variable on the fitting equation if the terminal velocity of the object.

**PRE LAB QUESTIONS**

1) Describe the difference between instantaneous and average acceleration?
2) Describe gravitational acceleration?
3) Describe free-falling of an object?
4) What are the possible forces acts on free-falling object?
5) Describe the drag force and terminal velocity?

## DATA ANALYSIS AND CALCULATIONS

*Data tables, calculations and graphs must be done in excel. All the excel functions should be added below each table in the lab report.*

### *Part A Gravitational acceleration by using free falling object*

Table 1 Calculation of average velocity and gravitational acceleration.

| $y_i$ [ ] | $t_i$ [ ] | $v = \frac{2y_i}{t_i}$ [ ] | $g_{cal} = \frac{v_{i+1} - v_i}{t_{i+1} - t_i}$ [ ] | PE $g_{cal}$ and $g$ = -9.81 m/s$^2$ |
|---|---|---|---|---|
| | | | | |
| | | | | |
| | | | | |
| | | | | |
| | | | | |
| | | | | |
| | | | | |
| | | | | |
| | | | | |
| | | | | |
| | | | | |
| | | | | |
| | | | | |
| | | | | |

- Make a graph of displacement $y_i$ vs time $t_i$.
- Do the quadratic fitting and calculate the gravitational constant by using the factor in front of the quadratic term. Then, compare (find percentage error) with $g$ = -9.81 m/s$^2$.
- Then, make another graph of velocity $v_i$ vs time $t_i$.
- Do the linear fitting and find the gravitational constant by using the slope of the graph. Then, compare (find percentage error) it with $g$ = -9.81 m/s$^2$.

***Part B Gravitational acceleration by using free falling object***

Table 2 Calculation of average velocity and gravitational acceleration.

| $y_i$ [ ] | $t_i$ [ ] | $v = \frac{2y_i}{t_i}$ [ ] | $g_{cal} = \frac{v_{i+1} - v_i}{t_{i+1} - t_i}$ [ ] | PE $g_{cal}$ and $g$ = -9.81 m/s$^2$ |
|---|---|---|---|---|
| | | | | |
| | | | | |
| | | | | |
| | | | | |
| | | | | |
| | | | | |
| | | | | |
| | | | | |
| | | | | |
| | | | | |
| | | | | |
| | | | | |
| | | | | |
| | | | | |

- Make a graph of $y_i$ vs time $t_i$. and include error bars.
- Do the quadratic fitting and calculate the gravitational constant by using the factor in front of the quadratic term. Then, compare (find percentage error) with $g$ = -9.81 m/s$^2$.
- Then, make another graph of $v_i$ vs time $t_i$. (include error bars).
- Do the linear fitting and find the gravitational constant by using the slope of the graph. Then, compare (find percentage error) it with $g$ = -9.81 m/s$^2$.

***Part C Terminal velocity of free-falling object***

- Explain the behavior of velocity of the object.
- Find the terminal velocity of the object by using y vs t data of the last 3 ft.
- Discuss the object motion and explain why the object velocity becomes constant?

# EXPERIMENT 5 PROJECTILE MOTION

## OBJECTIVE

Two-dimensional motion of an object is analyzed by using simultaneous vertical and horizontal axis motion of an object. Three types of projectile motions are analyzed, a) object moves from a height h with horizontal velocity, b) object moves from ground level with inclined to horizontal axis, and c) object moves from height h with inclined to horizontal axis. Initial velocity for part-a and maximum range for part-b are calculated. Gravitational acceleration is calculated by using part-c. Analytical results are compared with graphical results.

## THEORY AND PHYSICAL PRINCIPLES

In general, two-dimensional motion of an object under gravity is considered as the projectile motion and it is studied on the xy plane with respect to standard cartesian (xyz) coordinate system, which is also referred as the right-hand-coordinate system.

Objects on projectile motion can be analyzed by applying 1-dimensional kinematic equations on x axis and y axis as follows.

Following 3 linear equations can be used to study the y axis motion of projectile motion. Following variables are used; $v_0$ - initial velocity, v – velocity at time t, x or Δx – horizontal displacement, y or Δy – vertical displacement, $\theta$ - angle from horizontal axis, t - time difference from initial to final, $a$ – acceleration, $g$ – gravitational acceleration $g$ = 9.81 m/s$^2$

$$v_y = v_{0y} + gt \tag{1}$$

$$v_y^2 = v_{0y}^2 + 2g\,\Delta y \tag{2}$$

$$\Delta y = v_{0y}t + \frac{gt^2}{2} \tag{3}$$

Consider an object projected with only horizontal velocity $v_o$ from height $y$ above the ground. Time of the projectile can be found by using y-axis motion.

$$Y = \frac{gt^2}{2} \rightarrow t = \sqrt{\frac{2Y}{g}} \tag{4}$$

Then range can be measured experimentally. And the initial velocity can be found as follows:

$$v_0 = \frac{X}{t} \tag{5}$$

Consider an object projected inclined to the x axis from ground level. By measuring the range of projectile time of flight can be found.

$$t = \frac{X}{v_{ox}} = \frac{X}{v_o cos\theta} \tag{6}$$

By using time of flight gravitational acceleration can be calculated.

$$g = \frac{v_y - v_{0y}}{t} = \frac{-v_0^2\,2\,sin\theta cos\theta}{X} = \frac{-v_0^2\,sin2\theta}{X} \tag{7}$$

Gravitational acceleration ($g$) also can be found by using the slope of graph of rage (X) vs $sin2\theta$.

$$g = \frac{-v_0^2}{slope} \tag{8}$$

Initial angle to needed to get the maximum range (X) can be found by using graph of range (X) vs $\theta$. Which is a parabolic graph and maximum of parabola can be found by using the following equation. Parabolic fitting of graphs should be done first.

$$X = a\theta^2 + b\theta + c \tag{9}$$

$$\theta_{for\ max\ range} = \frac{-b}{2a} \tag{10}$$

Then, consider an object projected inclined to X axis and height-Y above the ground level with the same initial velocity and range will be measured experimentally.

Time of projectile: $$t = \frac{X}{v_{0x}} = \frac{X}{v_0 cos\theta} \tag{11}$$

By using the time of projectile and the vertical height, gravitational constant $g$ can be calculated.

$$Y = v_{0y}t + \frac{gt^2}{2} \rightarrow g = \frac{2(Y - t\, v_0\, sin\theta)}{t^2} \tag{12}$$

**APPARATUS AND PROCEDURE**

- This simulation will run on a web browser (Chrome or Firefox is better) and if you prefer then there is an option to download the java-applet. Here is the direct link to projectile simulation. https://phet.colorado.edu/sims/html/projectile-motion/latest/projectile-motion_en.html
- Here is a video link which shows details of data collection with simulation and data analysis (analytical and graphical) with excel. https://youtu.be/-gO8wFegkbE

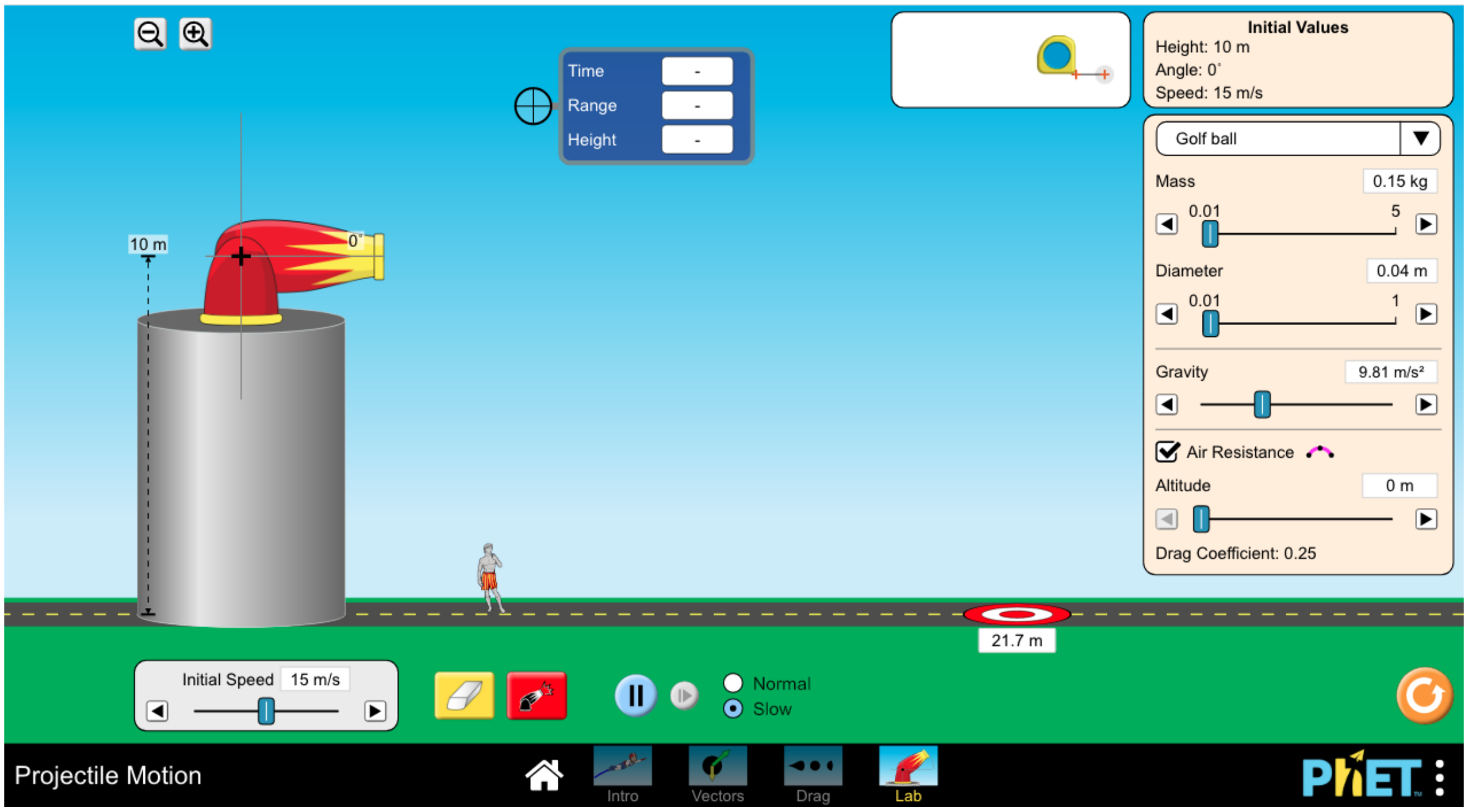

Figure 1 Simulator for projectile lab (Photo credit to https://phet.colorado.edu)

***Part A Horizontal projectile motion***

- Set the projectile launcher 15.0m above the ground level.
- Set the projectile inclined angle (from x axis) to zero.
- Set the initial velocity of the projectile to 14.0 m/s.
- Set the object to a golf ball (keep the mass, radius, and other options as default values).
- Click the red color fire button at the bottom of the simulator.
- Projectile path shows up in blue color. Click and drag the blue color box icon on top of the simulator. Move it to the end of the projectile path. Measure the time of flight and range of the projectile.
- Repeat the last two steps by decreasing height of the projectile 1.0m at a time.
- Calculate the initial velocity by using height and the range of the projectile on table 1.
- Find the average $v_{cal}$?
- Find the percent error between $v_{cal}$(avg) and v(given)?
- Make a graph of height (Y) vs $X^2$?
- Fit the data (trendline) with linear?
- Find the initial velocity ($v_{graph}$) by using slope of the graph?
- Find the percent difference between $v_{cal}$(graph) and v(given)?

***Part B Projectile from ground level***

- Set the projectile launcher height to ground level (zero meters).
- Set the launching angle to $25^0$ and initial velocity to 14 m/s.
- Click the red color fire button at the bottom of the simulator.
- Projectile path shows up in blue color. Click and drag the blue color box icon on top of the simulator. Move it to the end of the projectile path. Measure the time of flight and range of the projectile.
- Repeat the last two steps by increasing the initial launching angle of projectile $5.0^0$ at a time.
- Find the time of the projectile by using horizontal displacement of the projectile or range.
- Calculate the $g_{cal}$ by using time of flight, which should be calculated by using the range and the initial velocity.
- Calculate percent error between $g_{cal}$ and $g$ (-9.81 m/s$^2$)?
- Find the average $g_{cal}$(avg)?
- Find the percent error between $g_{cal}$(avg) and $g$?
- Make a graph of $\Delta x$ vs $sin2\theta$ ?
- Do the linear fitting to the graph and find the gravitational constant ($g_{graph}$)?
- Find the percent error between $g_{graph}$ and $g$?
- Then, make a graph of X (on y-axis of the graph) vs theta (on x-axis of the graph).
- Fit the data with polynomial fitting.
- Find the theta when range is maximum $\theta_{max}$?
- Maximum value of parabolic function can be done by using [-(b/2a)].

***Part C Projectile above ground level with initial launching angle***

- Set the projectile launcher height to 3.0m above ground level.
- Set the launching angle to $10^0$ and initial velocity to 14 m/s.
- Click the red color fire button at the bottom of the simulator.
- Projectile path shows up in blue color. Click and drag the blue color box icon on top of the simulator. Move it to the end of the projectile path. Measure the time of flight and range of the projectile.
- Repeat the last two steps by increasing the initial launching angle of projectile $10^0$ at a time.
- Calculate the $g_{cal}$ by using time of flight, which should be calculated by using the range and the initial velocity.
- Calculate percent error between $g_{cal}$ and $g$ (-9.81 m/s$^2$)?
- Find the average $g_{cal}$(avg)?
- Find the percent error between $g_{cal}$(avg) and $g$?

**PRE LAB QUESTIONS**

1) Consider an object sends into projectile and describe y axis motion of the object?
2) Consider an object sends into projectile and describe x axis motion of the object?
3) Consider an object sends into projectile from ground level with initial velocity of $v_0$ and $\theta$ angle from horizontal axis.
    a) Find and equation for time to reach the maximum height of the projectile?
    b) Find and equation for maximum height of the projectile?
    c) Find and equation for range the projectile?
4) Angle to get maximum range is about 45 degrees when an object is sent from ground level. Is it the same when an object sends to projectile h distance above the ground? Explain your answer why or why not?

## DATA TABLES AND CALCULATIONS

*Data tables, calculations and graphs must be done in excel. All the excel functions should be added below each table in the lab report.*

### *Part A Horizontal projectile motion*

Table 1 Analysis of horizontal projectile motion

| Vertical Height Y [ ] | Horizontal Displacement X [ ] | $X^2$ [ ] | Initial velocity calculated $v_{cal}$ [ ] |
|---|---|---|---|
| | | | |
| | | | |
| | | | |
| | | | |
| | | | |
| | | | |
| | | | |
| | | | |
| | | | |
| | | | |
| | | | |
| | | | |

- Find the average $v_{cal}$?
- Find the percent error between $v_{cal}$(avg) and v(given)?
- Make a graph of height (Y) vs $X^2$?
- Fit the data (trendline) with linear?
- Find the initial velocity ($v_{graph}$) by using slope of the graph?
- Find the percent difference between $v_{cal}$(graph) and v(given)?

### *Part B Projectile from ground level*

- Find the time of the projectile by using horizontal displacement of the projectile or range.
- Calculate the $g_{cal}$ by using time of flight, which should be calculated by using the range and the initial velocity.
- Calculate percent error between $g_{cal}$ and $g$(-9.81 m/s$^2$)?
- Find the average $g_{cal}$(avg)?
- Find the percent error between $g_{cal}$(avg) and $g$?
- Make a graph of $\Delta x$ vs $sin2\theta$ ?
- Do the linear fitting to the graph and find the gravitational constant ($g_{graph}$)?
- Find the percent error between $g_{graph}$ and $g$?

- Then, make a graph of X (on y-axis of the graph) vs theta (on x-axis of the graph).
- Fit the data with polynomial fitting.
- Find the theta when range is maximum $\theta_{max}$?
- Maximum value of parabolic function can be done by using [-(b/2a)].

Table 2 Measurements of launching angle vs range and calculation of gravitational acceleration

| Initial Angle $\theta$ [ ] | X [ ] | $t_{cal} = \frac{\Delta x}{v_{0x}}$ ($v_{0x} = v_0 cos\theta$) [ ] | $g_{cal} = \frac{-v_0^2 sin2\theta}{\Delta x}$ [ ] | *PE between $g_{cal}$ and -9.81m/s²* |
|---|---|---|---|---|
| 25.0 | | | | |
| 30.0 | | | | |
| 35.0 | | | | |
| 40.0 | | | | |
| 45.0 | | | | |
| 50.0 | | | | |
| 55.0 | | | | |
| 60.0 | | | | |
| 65.0 | | | | |
| 70.0 | | | | |

***Part C Projectile above ground level with initial launching angle***

- Set the height of the projectile to 3.0m above the ground.
- Find the time of the projectile by using horizontal displacement of the projectile or range.
- Calculate the $g_{cal}$ by using time of flight, which should be calculated by using the range and the initial velocity.
- Calculate percent error between $g_{cal}$ and $g$ (-9.81 m/s$^2$)?
- Find the average $g_{cal}$(avg)?
- Find the percent error between $g_{cal}$(avg) and $g$?

Table 3 Measurements of launching angle vs range and calculation of gravitational acceleration

| Initial Angle $\theta$ [ ] | X [ ] | $t_{cal} = \frac{X}{v_{0x}}$ ($v_{0x} = v_0 cos\theta$) [ ] | $g_{cal} = \frac{2(-Y - v_{0y}t_{cal})}{t_{cal}^2}$ ($v_{0y} = v_0 sin\theta$) [ ] | *PE between* $g_{cal}$ *and* $-9.81 m/s^2$ |
|---|---|---|---|---|
| 10.0 | | | | |
| 20.0 | | | | |
| 30.0 | | | | |
| 35.0 | | | | |
| 40.0 | | | | |
| 45.0 | | | | |
| 50.0 | | | | |
| 60.0 | | | | |

# EXPERIMENT 6 NEWTON LAWS AND MECHANICAL ENERGY

## OBJECTIVE

- To understand Newton's second law and mechanical energy.
- Use Newton's second law to calculate the acceleration and force acting on moving objects.
- Measure kinetic energy of moving objects and calculate potential energy of an object.
- Compare the changing kinetic energy and changing potential energy object.

## THEORY AND PHYSICAL PRINCIPLES

When an object is moving on linear air track without any friction, consider all the forces acting on the object and use Newton's second law to find acceleration and the next external force acts on the object.

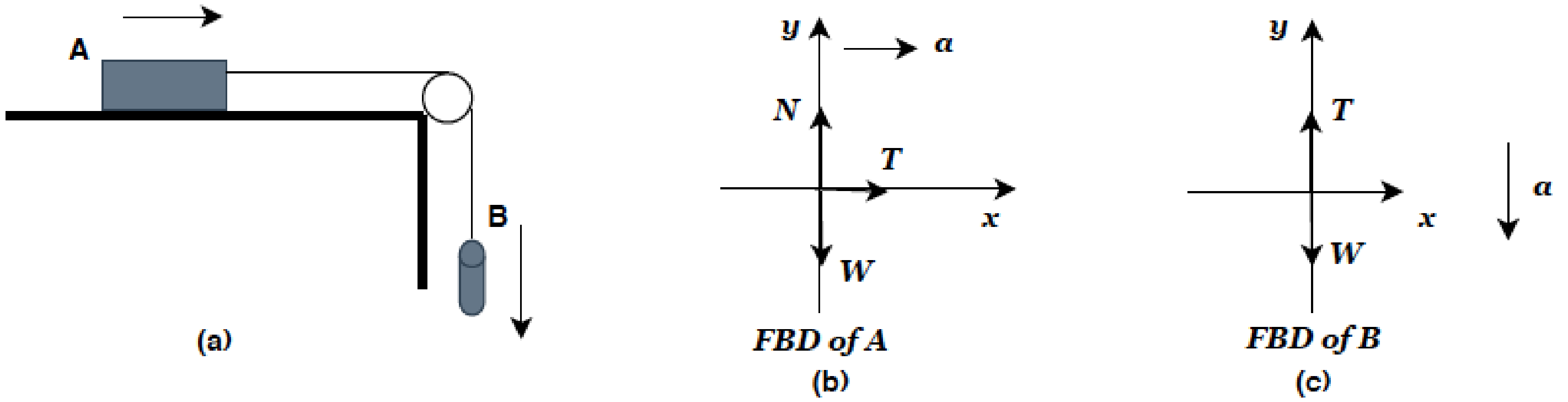

Figure 1 (a) Object moving on linear air track, (b) Free-body-diagram (FBD) of object-A and (c) Free-body-diagram (FBD) of object-B

In figure 1(a) there are two objects attached together with a string (non-stretchable). When object B is released from the rest, h distance above the ground, object A and B both are accelerating at the same magnitude of $a$.

### *Part A Newton's second law*

By applying Newton's second law to object A and B, acceleration and tension can be found. Since the object A is on the air-track, there is no friction.

Object A: $T = M_2\, a$ (1)

$\mathrm{N} - \mathrm{W} = 0 \rightarrow N = M_2 g$ (2)

Object B: $T - M_1\, g = -M_1 a$ (3)

By solving equations (1) and (3), tension can be found:

$$T = \left(\frac{M_1 M_2}{M_1 + M_2}\right) g \quad (4)$$

$$a = \frac{M_1 g}{M_1 + M_2} \quad (5)$$

***Part B Mechanical Energy***

Consider that object A starts with zero velocity on x-axis and object B also starts at zero velocity on y-axis and it is h distance above the ground level.

Total energy of the system when object start at rest:

$$E_{tot}(initial) = KE_i + PE_i = M_1 g h_1 \quad (6)$$

Since both objects are at rest at the beginning the kinetic energy of the system is zero.

Total energy of the system at the end of motion (end is when the object B reach the ground):

$$E_{tot}(final) = KE_i + PE_i = \frac{1}{2}(M_1 + M_2)v^2 + M_1 g h_2 \quad (7)$$

At final, both objects are moving with velocity v and there is no potential energy of the system.

If there is no energy lost during the motion, the final total energy of the system should be the same as the initial total energy of the system.

Final velocity of the objects can be calculated by using equations (6) and (7).

$$v = \sqrt{\frac{2M_1 g(h_1 - h_2)}{M_1 + M_2}} = \sqrt{\frac{2M_1 g(\Delta h)}{M_1 + M_2}} \quad (8)$$

***Part C Newton 2nd law with friction***

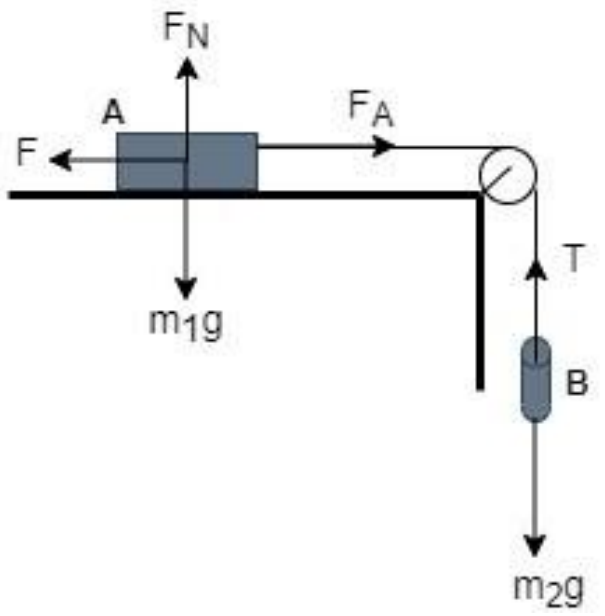

Figure 2 Object moving on linear air track with friction.

When friction is present then the related equations for acceleration and tension the system change.

Object A: $$T - F = m_2\, a \quad (9)$$

$$F = \mu N = \mu m_2 g \quad (10)$$

Object B:

$$T - m_1\, g = -m_1 a \quad (11)$$

By solving equations (1) and (3), tension can be found:

$$T = m_1 g - m_1 a = \frac{(\mu+1)m_1 m_2 g}{m_1 + m_2} \quad (12)$$

$$a = \frac{m_1 g - \mu m_2 g}{m_1 + m_2} \quad (13)$$

## APPARATUS AND PROCEDURE

- This experiment is done with following simulation: https://ophysics.com/f3.html
- This simulation works on any web browser (Chrome or Firefox is recommended.)
- A very detail video tutorial with data collection from simulator and data analysis with excel can be found here: https://youtu.be/mgTImjdHhsw

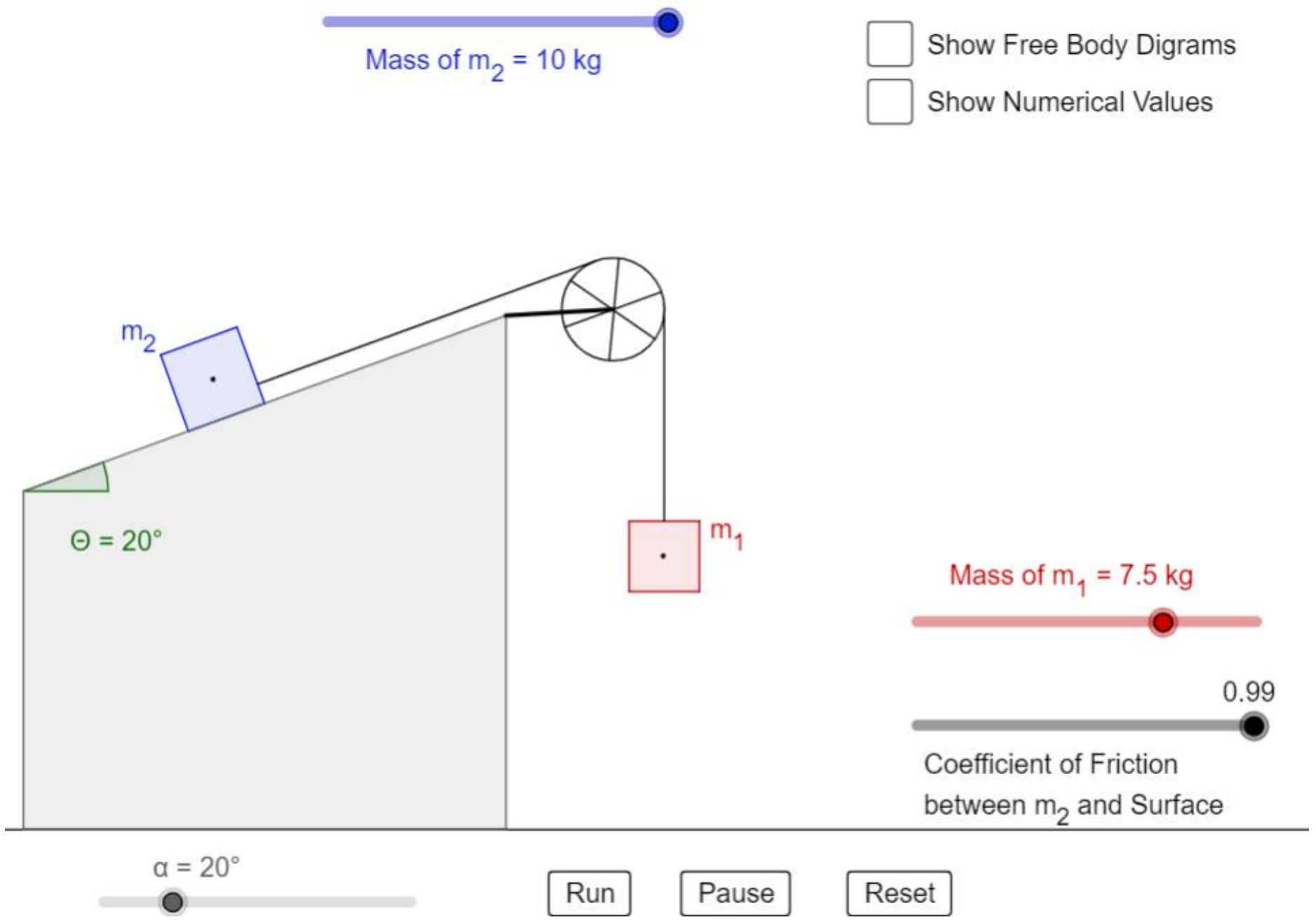

Figure 3 Simulation for Newton's second law experiment (Picture credit, https://ophysics.com/)

### *Part A Newton's second law (without friction force)*

- Set the inclined angle and the friction coefficient of the horizontal plane to zero.
- Set the mass of the object on the horizontal plane to $M_2$=10.00 kg. Use two decimal place accuracy of the mass of the object 2.
- Set the mass of the hanger $M_1$=1.00 kg. Use two decimal place accuracy of the mass of the object 1.
- Click on "show numerical values" on top right of the simulator.
- Click on "run" on the bottom of the simulator. After run is done collect the following information, vertical height (h) of the mass $M_1$, time of motion (t).
- Change the mass of the hanger by 1.00 kg at a time and repeat the last step five more times.
- Calculate final velocity and the acceleration by using a linear one dimensional equation?
- Calculate the acceleration of the system by using Newton laws?
- Compare it with the acceleration from the linear equation?
- Calculate the tension of the system by using acceleration from linear equations and also by using calculated acceleration by using Newton laws?
- Compare the tensions of the system calculated above in two different methods?
- Make a graph of tension vs acceleration (with experimentally calculated values) for the object $M_2$?
- Fit the data with a linear trend line and find the mass $M_2$ by using slope of the graph?
- Compare the $M_2$ from the graph with the known value of $M_2$?

***Part B Mechanical Energy***

- Energy of the system can be calculated by knowing the mass of the objects ($M_1$ and $M_2$), initial height (h) of the mass $M_1$ and final velocity (v) of the system.
- Calculate initial total energy of the system?
- Calculate final total energy of the system?
- Check the law of conservation energy?
- Make a graph of KE(final) vs PE(initial).
- Do the linear fitting and find the slope of the graph?
- Compare slope of the graph with expected value to investigate the law of conservation of energy?

***Part C Newton 2nd law with friction***

- Set the inclined angle to zero.
- Set the friction coefficient of the horizontal plane to 0.14.
- Set the mass of the object on the horizontal plane to $M_2$=10.00 kg. Use two decimal place accuracy of the mass of the object 2.
- Set the mass of the hanger $M_1$=3.00 kg. Use two decimal place accuracy of the mass of the object 1.
- Click on “show numerical values” on top right of the simulator.
- Click on “run” on the bottom of the simulator. After run is done collect the following information, vertical height (h) of the mass $M_1$, time of motion (t).
- Change the mass of the hanger by 1.00 kg at a time and repeat the last step five more times.
- Calculate final velocity and the acceleration by using a linear one-dimensional equation?
- Calculated the acceleration of the system by using Newton laws?
- Compare it with the acceleration from linear equations?
- Calculate the tension of the system by using acceleration from linear equations and also by using calculated acceleration by using Newton laws?
- Compare the tensions of the system by using acceleration calculated above in two different methods?
- Make a graph of tension vs acceleration (with experimentally calculated values) for the object $M_2$?
- Fit the data with a linear trend line and find the mass $M_2$ by using slope of the graph and also by using y intercept?
- Compare the $M_2$ values from the graph with known values of $M_2$?
- Calculate the energy loss of the system?
- Make a graph of KE(final) vs PE(initial).
- Do the linear fitting and find the slope of the graph?
- Compare slope of the graph with expected value to investigate the law of conservation of energy?
- Find the total energy loss due to friction force by using y intercept of the graph?

**PER LAB QUESTIONS**

1. Describe Newton's laws?
2. Describe friction force?
3. Describe all the possible forces acting on an object moving on a horizontal surface?
4. Describe mechanical energy?
5. How would the conservation of energy affect in the presence of friction?

## DATA ANALYSIS AND CALCULATIONS

*Data tables, calculations and graphs must be done in excel. All the excel functions should be added below each table in the lab report.*

### *Part A Newton's second law (without friction force)*

Table 1 Analysis of acceleration of the system by using Newton laws

| Hanger Mass $M_1$ [ ] | Height h [ ] | Time t [ ] | Final velocity, v [ ] | $a$_exp [ ] | $a$_cal [ ] | *PD of accelerations* |
|---|---|---|---|---|---|---|
| | | | | | | |
| | | | | | | |
| | | | | | | |
| | | | | | | |
| | | | | | | |
| | | | | | | |

Table 2 Analysis of tension force by Newton laws

| Hanger Mass $M_1$ [ ] | $T$_exp [ ] | $T$_cal [ ] | *PD of Tensions* |
|---|---|---|---|
| | | | |
| | | | |
| | | | |
| | | | |
| | | | |
| | | | |

- Make a graph of tension vs acceleration (with experimentally calculated values) for the object $M_2$?
- Fit the data with a linear trend line and find the mass $M_2$ by using slope of the graph?
- Compare the $M_2$ from the graph with the known value of $M_2$?

***Part B Mechanical Energy***

Table 3 Mechanical energy of the system

| Hanger Mass $M_1$ [ ] | $E_{tot}(final) =$ KE_final [ ] | $E_{tot}(initial) =$ PE_initial [ ] | *PD of $E_{tot}(initial)$ and $E_{tot}(final)$* |
|---|---|---|---|
| | | | |
| | | | |
| | | | |
| | | | |
| | | | |
| | | | |

- Make a graph of KE(final) vs PE(initial).
- Do the linear fitting and find the slope of the graph?
- Compare slope of the graph with expected value to investigate the law of conservation of energy?

***Part C Newton 2nd law with friction***

Table 4 Analysis of acceleration of the system by Newton laws with friction

| Hanger Mass $M_1$ [ ] | Height h [ ] | Time t [ ] | Final velocity, v [ ] | $a$_exp [ ] | $a$_cal [ ] | *PD of accelerations* |
|---|---|---|---|---|---|---|
| | | | | | | |
| | | | | | | |
| | | | | | | |
| | | | | | | |
| | | | | | | |
| | | | | | | |

Table 5 Analysis of tension force by Newton laws with friction

| Hanger Mass $M_1$ [ ] | $T$_exp [ ] | $T$_cal [ ] | *PD of Tensions* |
|---|---|---|---|
| | | | |
| | | | |
| | | | |
| | | | |
| | | | |
| | | | |

- Make a graph of tension vs acceleration (with experimentally calculated values) for the object $M_2$?
- Fit the data with a linear trend line and find the mass $M_2$ by using slope of the graph and also by using y intercept?
- Compare the $M_2$ values from the graph with known values of $M_2$?

### *Energy loss due to friction*

Table 6 Mechanical energy loss due to friction

| Hanger Mass $M_1$ [ ] | $E_{tot}(final) =$ KE_final [ ] | $E_{tot}(initial) =$ PE_initial [ ] | *Energy loss due to friction* $\Delta E$ [ ] |
|---|---|---|---|
| | | | |
| | | | |
| | | | |
| | | | |
| | | | |
| | | | |

- Make a graph of KE(final) vs PE(initial).
- Do the linear fitting and find the slope of the graph?
- Compare slope of the graph with expected value to investigate the law of conservation of energy?
- Find the total energy loss due to friction force by using y intercept of the graph?

# EXPERIMENT 7 FRICTION FORCE

## OBJECTIVE

Goal is to calculate both static and kinetic friction coefficients between a given object and platform surface. Different methods to calculate the friction coefficients are investigated. Correlation of inclined angle and friction coefficients is investigated.

## THEORY AND PHYSICAL PRINCIPLES

### *Part A: Object on a level plane*

For a variety of surfaces the ratio of the frictional force (*F, static or kinetic*) to the normal force ($F_N$) is approximately constant. This ratio defines the coefficient of friction:

$$\mu = \frac{F}{F_N} \tag{1}$$

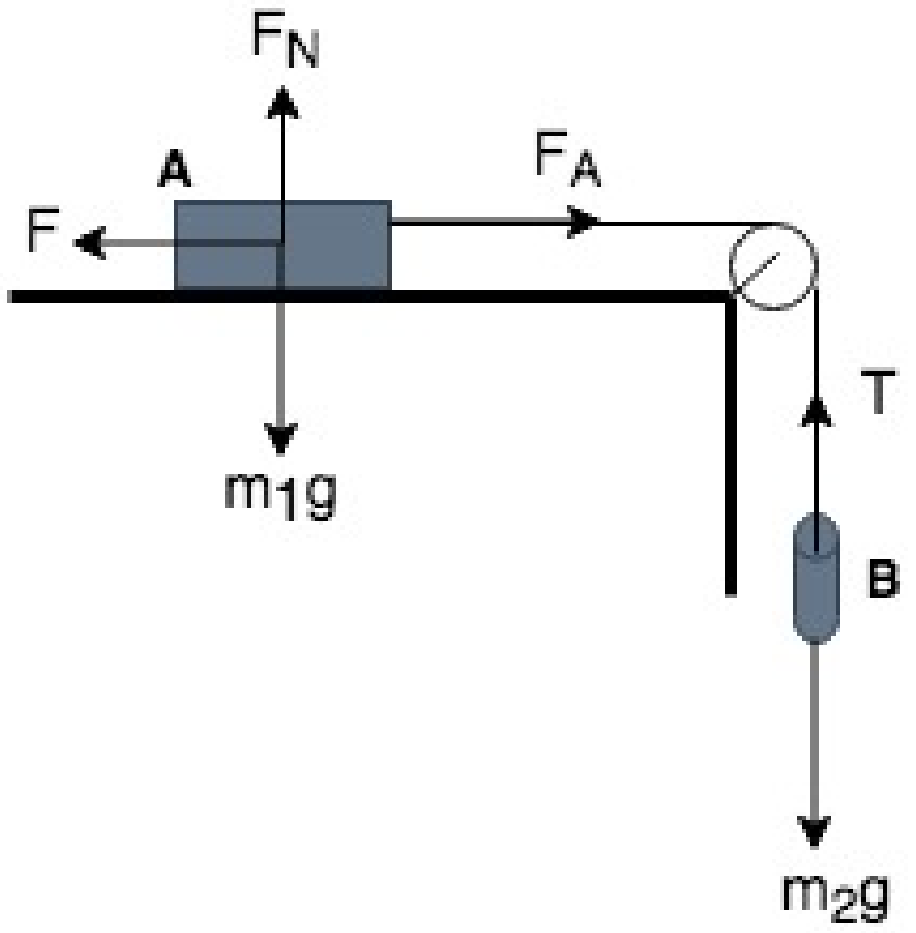

Figure 1 Schematic diagram of the object and free-body-diagram

In the static case when the applied force reaches a value such that the object instantaneously starts to move we obtain the maximum frictional force or limiting value of the frictional force $F_{max}$.

We can now obtain the coefficient of static friction coefficient: $\mu_s = \frac{F_{max}}{F_N}$ (2)

When the object is moving it experiences a frictional force, $F_K$ that is less than the static friction force. Frictional experiments tell us that we can (analogous to the static case) define a coefficient of kinetic friction given by:

$$\mu_k = \frac{F_k}{F_N} \tag{3}$$

Considering the Figure 1 above and applying Newton's 2nd law:

$$F_{max} = m_2 g \tag{4}$$

$$F_N = m_1 g \tag{5}$$

$$\mu = \frac{F_{max}}{F_N} = \frac{m_2}{m_1} \tag{6}$$

## *Part B: Object freely moves on an inclined plane*

Consider an object is at rest on the inclined plane as in Figure 2 and applying Newton's 2nd law:

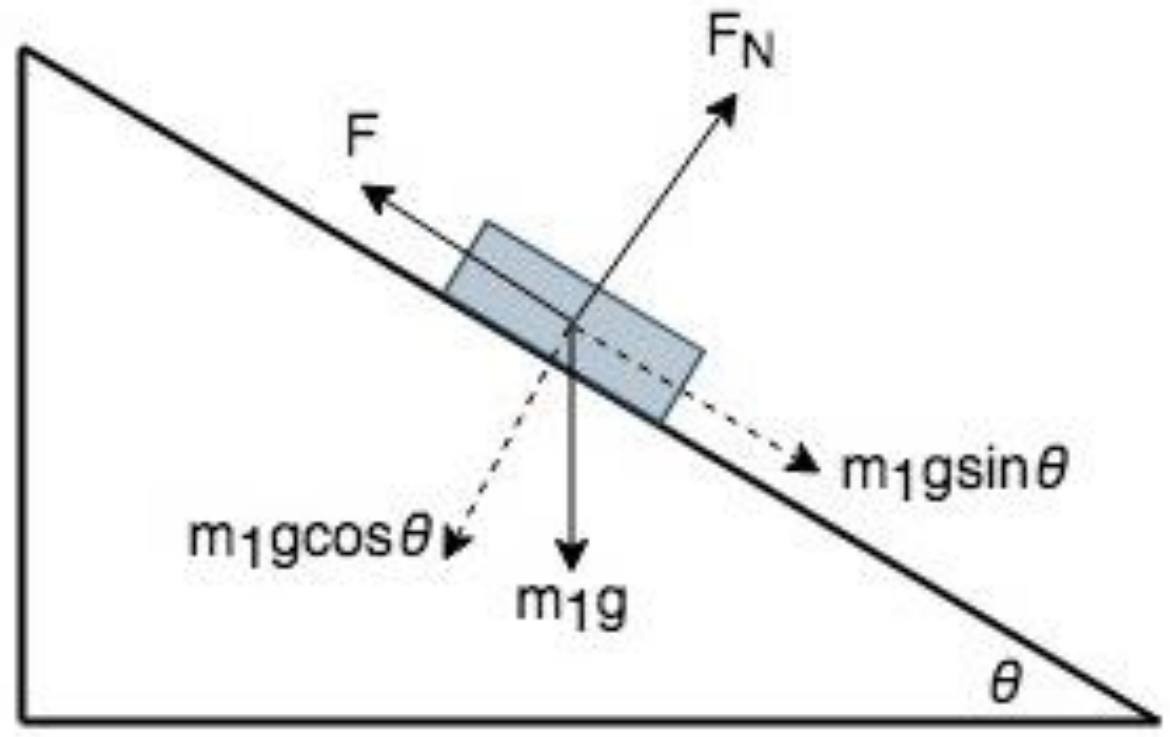

Figure 2 Schematic diagram of the object and free-body-diagram of object on inclined plane

$$\mu = \frac{F}{F_N} \tag{7}$$

$$F = m_1 g sin\theta \tag{8}$$

$$F_N = m_1 g cos\theta \tag{9}$$

$$\mu = tan\theta \tag{10}$$

## APPARATUS AND PROCEDURE

### Part A Static and Kinetic friction coefficients with force sensor

- This part of the experiment is done with the following simulation: https://www.thephysicsaviary.com/Physics/Programs/Labs/ForceFriction/
- A very detail video tutorial with data collection from simulator and data analysis with excel can be found here: https://youtu.be/r6kKn1cf9Gw

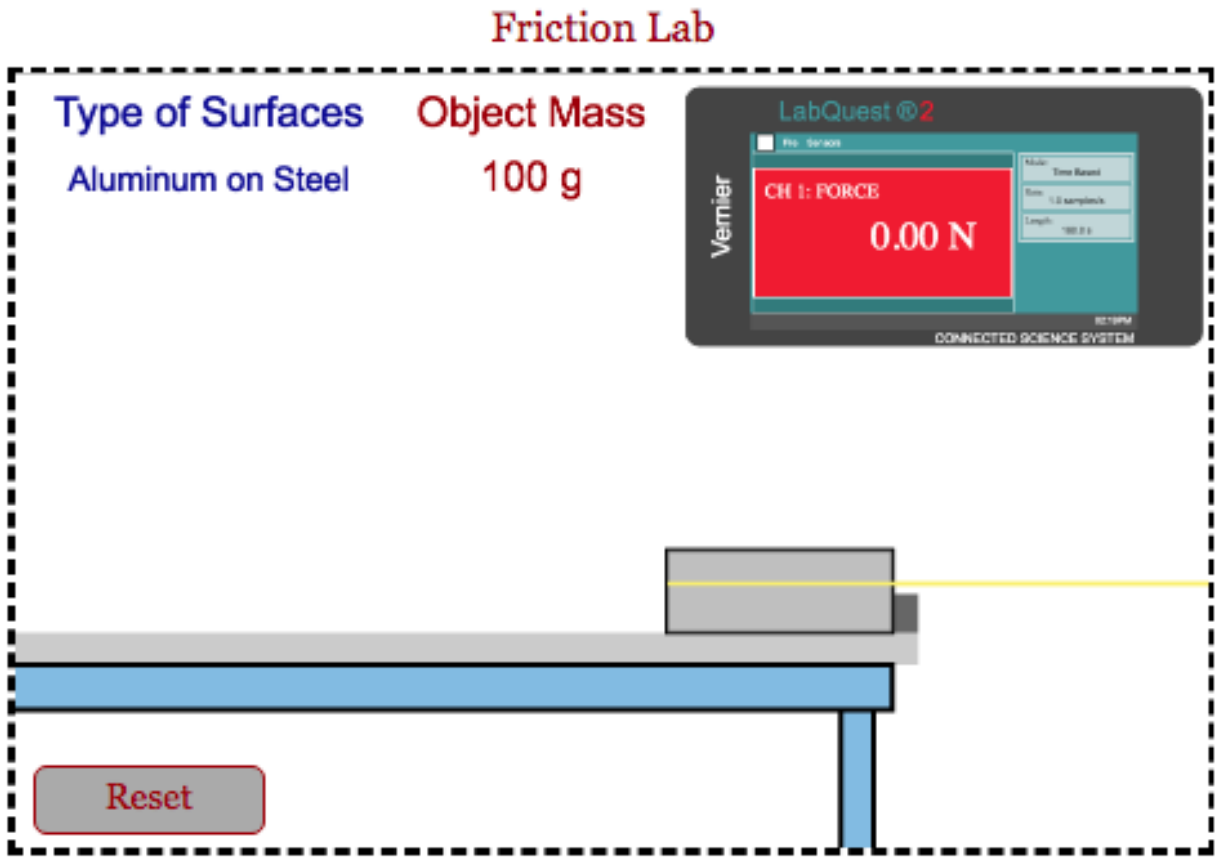

Figure 3 Simulation for friction experiment (Picture credit, https://www.thephysicsaviary.com/)

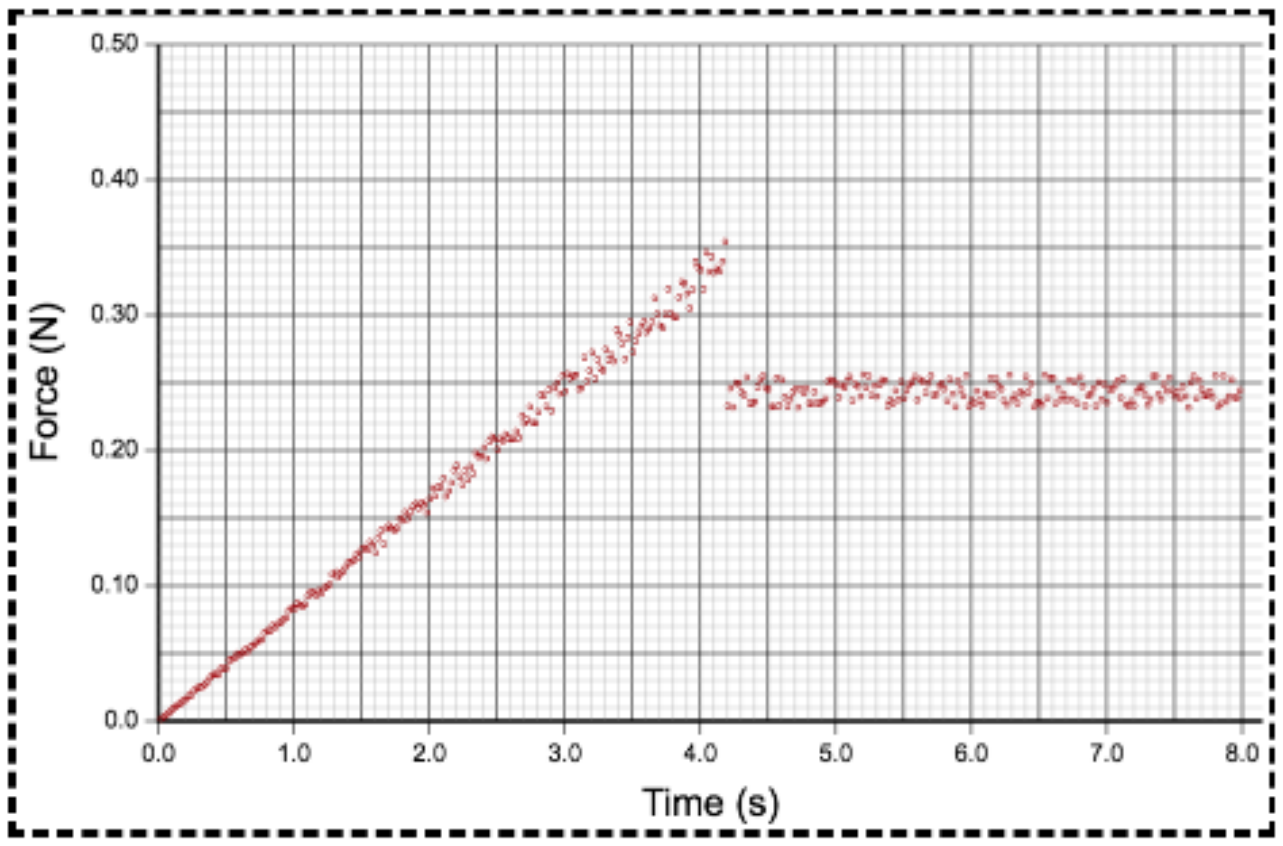

Figure 4 Simulation results of friction force as a function of time (Picture credit, https://www.thephysicsaviary.com/)

- Click on the second line of the blue color word on the top left corner to change the type of object and the surface.
- Select the “wood on table” by clicking on the blue color word on top left corner.
- Change the mass of the object (cart) by clicking the red arrows (up and down) in the middle top of the simulator.
- Change the mass to the lowest possible value of 100.00 grams to start with.
- Click on the start button on the bottom left side to start the experiment.
- When the start button is clicked a graph of friction force vs time appears on bottom of the simulator page which looks like the figure 5.
- By using the graph of friction force vs time, extract the maximum value of static friction force and the kinetic friction force.
- Then click the reset button on bottom left on the simulator.
- Repeat the process with increasing mass of the cart.
- Then repeat the procedure by changing the object and surface but keeping the same mass of the object.

**Part B Static friction coefficient by using hanger mass**

- Following simulation is used for this part of the experiment.
- http://amrita.olabs.edu.in/?sub=1&brch=5&sim=191&cnt=4

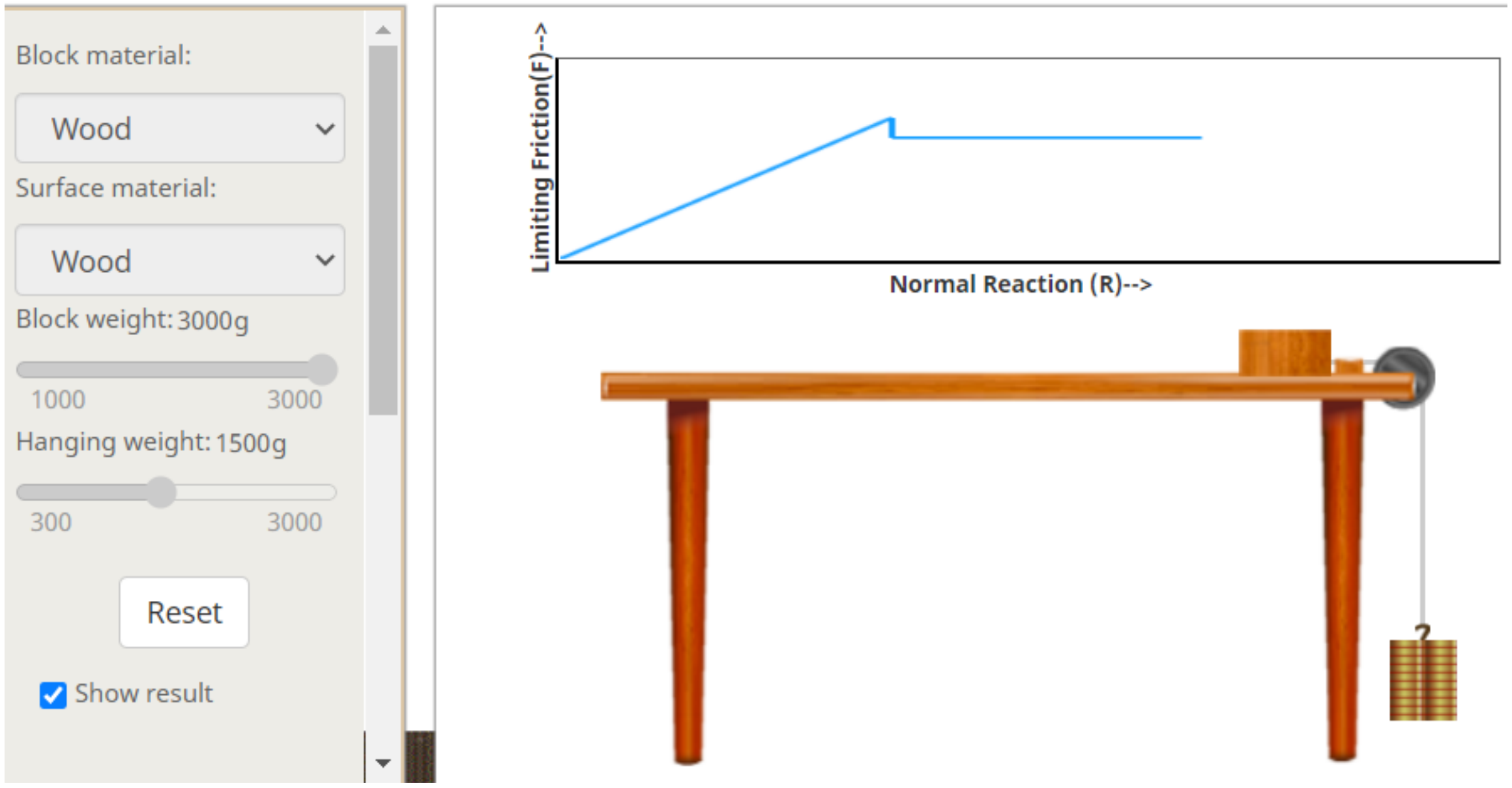

Figure 5 Schematic diagram of the object and free-body-diagram of object on inclined plane (Picture credit, http://amrita.olabs.edu.in)

- Set the block material on the top left corner to “wood”.
- Set the surface material on the top left corner to “wood”.
- Set the block weight to 1000.0 grams.
- Increase the hanger weight slowly (click on the slider then use the arrows on the keyboard) till the block on the table starts to move with constant velocity.
- Record weights of both blocks and the hanger weight.
- Repeat above procedure by increasing block weight.

## Part B Static friction coefficient by using inclined angle

- Following simulation can be used to find the static. Click the link to access the simulation.
- https://www.thephysicsaviary.com/Physics/Programs/Labs/ForcesOnInclineLab/index.html
- Change the block mass to lowest possible value by clicking the mass value on top right of the simulator.
- Then click on "click to change the angle" on bottom right. Simulator changes the angle of the platform and at a certain point the object will start to move downward on the inclined surface.
- When the object starts the motion click on "freeze" bottom right. And record the angle.
- Reset the simulator and repeat the last two steps by increasing the mass of the object by clicking mass value in grams on top right.

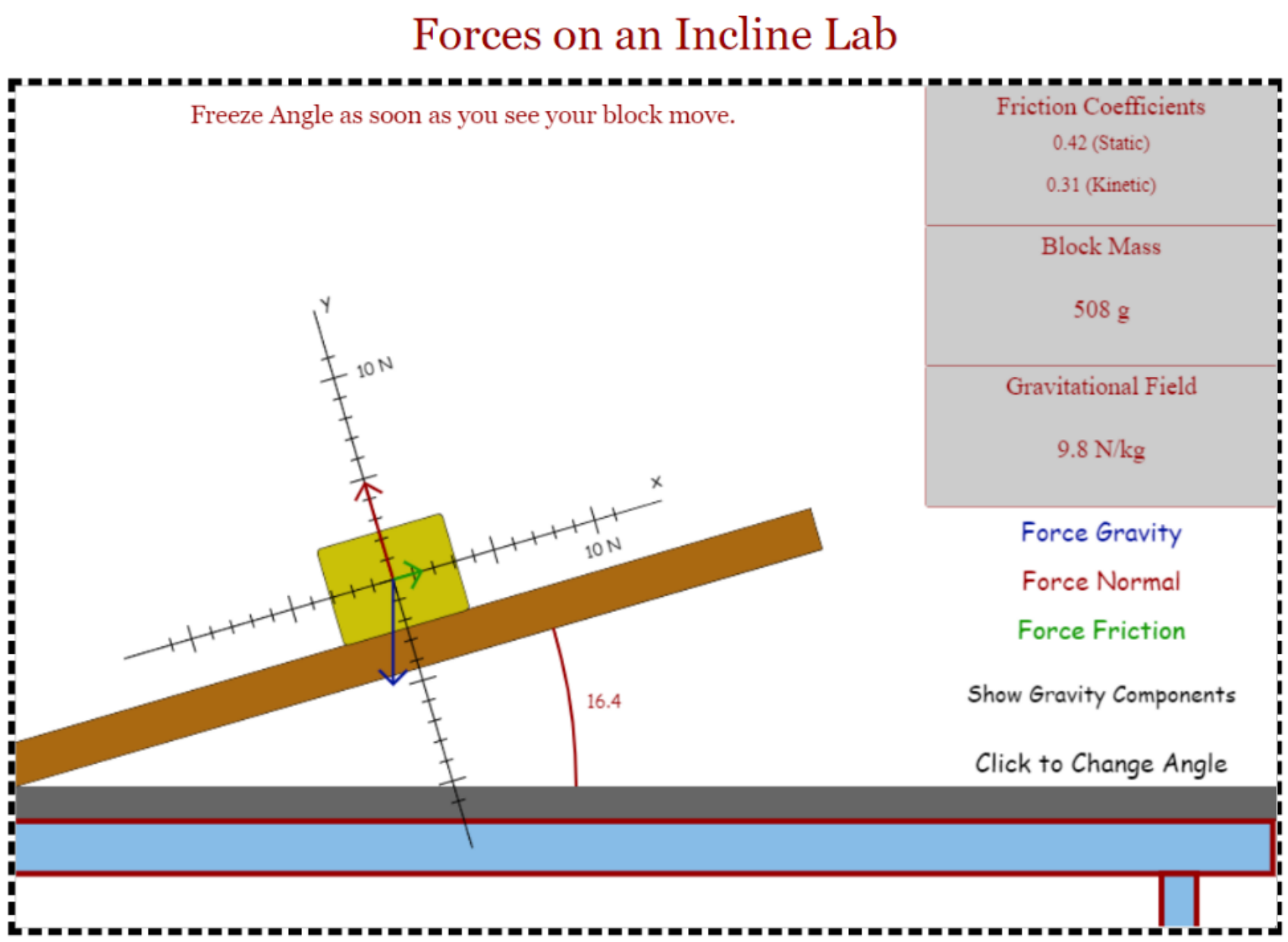

Figure 6 Schematic diagram of the object and free-body-diagram of object on inclined plane (Picture credit, https://www.thephysicsaviary.com)

## PRE LAB QUESTIONS

1) What is the friction force?
2) Describe static friction coefficient?
3) Describe kinetic friction coefficient?
4) What is the highest friction force?
5) Which coefficient, static or kinetic is usually the larger?
6) Is it possible to have a coefficient of friction greater than 1? Justify your answer.

## DATA ANALYSIS AND CALCULATIONS

*Data tables, calculations and graphs must be done in excel. All the excel functions should be added below each table in the lab report.*

### *Part A Friction coefficients-wood on lab table*

Table 1 Friction coefficients (static and kinetic) of wood on table

| Mass of the object M [ ] | Normal Force $F_N$ [ ] | Maximum static friction force, $F_s$ [ ] | Kinetic friction force, $F_k$ [ ] | Static friction coefficient $\mu_s(cal)$ | Kinetic friction coefficient $\mu_k(cal)$ |
|---|---|---|---|---|---|
| | | | | | |
| | | | | | |
| | | | | | |
| | | | | | |
| | | | | | |

- Find the average static friction coefficient? $\mu_s(cal_avg) =$
- Find the average kinetic friction coefficient? $\mu_k(cal_avg) =$
- Make a graph of $F_s$ vs $F_N$. Do the data fitting with the trend line (linear). Find the slope of the graph.
  Slope of the graph = $\mu_s(graph) =$
- Find the percent difference between $\mu_s(cal_avg)$ and $\mu_s(graph)$?
- Make a graph of $F_k$ vs $F_N$. Do the data fitting with the trend line (linear). Find the slope of the graph.
  Slope of the graph = $\mu_k(graph) =$
- Find the percent difference between $\mu_k(cal_avg)$ and $\mu_k(graph)$?

### *Friction coefficients-rubber on ice*

Table 2 Friction coefficients (static and kinetic) of rubber on ice

| Mass of the object M [ ] | Normal Force $F_N$ [ ] | Maximum static friction force, $F_s$ [ ] | Kinetic friction force, $F_k$ [ ] | Static friction coefficient $\mu_s(cal)$ | Kinetic friction coefficient $\mu_k(cal)$ |
|---|---|---|---|---|---|
| | | | | | |
| | | | | | |
| | | | | | |
| | | | | | |
| | | | | | |

- Find the average static friction coefficient? $\mu_s(cal_avg) =$
- Find the average kinetic friction coefficient? $\mu_k(cal_avg) =$
- Make a graph of $F_s$ vs $F_N$. Do the data fitting with the trend line (linear). Find the slope of the graph.
  Slope of the graph = $\mu_s(graph) =$
- Find the percent difference between $\mu_s(cal_avg)$ and $\mu_s(graph)$?
- Make a graph of $F_k$ vs $F_N$. Do the data fitting with the trend line (linear). Find the slope of the graph.
  Slope of the graph = $\mu_k(graph) =$
- Find the percent difference between $\mu_k(cal_avg)$ and $\mu_k(graph)$?

***Friction coefficients-aluminum on steel***

Table 3 Friction coefficients (static and kinetic) of aluminum on steel

| Mass of the object M [ ] | Normal Force $F_N$ [ ] | Maximum static friction force, $F_s$ [ ] | Kinetic friction force, $F_k$ [ ] | Static friction coefficient $\mu_s(cal)$ | Kinetic friction coefficient $\mu_k(cal)$ |
|---|---|---|---|---|---|
| | | | | | |
| | | | | | |
| | | | | | |
| | | | | | |
| | | | | | |

- Find the average static friction coefficient? $\mu_s(cal_avg) =$
- Find the average kinetic friction coefficient? $\mu_k(cal_avg) =$
- Make a graph of $F_s$ vs $F_N$. Do the data fitting with the trend line (linear). Find the slope of the graph.
  Slope of the graph = $\mu_s(graph) =$
- Find the percent difference between $\mu_s(cal_avg)$ and $\mu_s(graph)$?
- Make a graph of $F_k$ vs $F_N$. Do the data fitting with the trend line (linear). Find the slope of the graph.
  Slope of the graph = $\mu_k(graph) =$
- Find the percent difference between $\mu_k(cal_avg)$ and $\mu_k(graph)$?

***Friction coefficients of other object and surfaces***

Table 4 Friction coefficients (static and kinetic) with simulation

| Object and surface | Mass of the object M [ ] | Normal Force $F_N$ [ ] | Maximum static friction force, $F_s$ [ ] | Kinetic friction force, $F_k$ [ ] | Static friction coefficient $\mu_s(cal)$ | Kinetic friction coefficient $\mu_k(cal)$ |
|---|---|---|---|---|---|---|
| | | | | | | |
| | | | | | | |
| | | | | | | |

***Part B Static friction coefficient by using object and hanger mass system***

Table 5 Static friction coefficient by object and hanger mass system

| Mass of the object M [ ] | Normal Force $F_N$ [ ] | Maximum static friction force, $F_s$ [ ] | Static friction coefficient $\mu_s(cal)$ |
|---|---|---|---|
| | | | |
| | | | |
| | | | |
| | | | |
| | | | |

- Find the average static friction coefficient? $\mu_s(cal_avg) =$
- Make a graph of $F_s$ vs $F_N$. Do the data fitting with the trend line (linear). Find the slope of the graph.
  Slope of the graph = $\mu_s(graph) =$
- Find the percent difference between $\mu_s(cal_avg)$ and $\mu_s(graph)$?

**Part B Static friction efficient by using inclined angle**

Table 6 Static friction coefficient by using inclined plane

| Mass of the object M [ ] | Angle when object start moving θ [ ] | Static friction coefficient $\mu_s(cal)$ | PE $\mu_s(cal)$ and $\mu_s(known)$ |
|---|---|---|---|
| | | | |
| | | | |
| | | | |
| | | | |
| | | | |

# EXPERIMENT 8 CENTRIPETAL FORCE

## OBJECTIVE

Through this experiment you will understand the centripetal force of an object in uniform circular motion. You will also understand how Newton's second law can be used to find a conceptual connection between physical force and centripetal force.

## THEORY AND PHYSICAL PRINCIPLES

When an object moving in a circular orbit of radius $(r)$ with constant angular velocity $(\omega)$, magnitude of tangential velocity $(v)$ is constant.

$$v = r\omega \quad (1)$$

The period $(T)$ of circular motion is generally connected to frequency of the motion,

$$f = \frac{1}{T} \quad (2)$$

Tangential velocity can be found by the total linear distance traveled in complete circle (circumference of the orbit) and the time per one complete circle (period).

$$v = \frac{2\pi r}{T} \quad (3)$$

$$v = r2\pi f = r\omega \quad (4)$$

$$\omega = \frac{2\pi}{T} = 2\pi f \quad (5)$$

However, direction of the tangential velocity changes every second on circular orbit, because of that a linear acceleration produce towards the center of the orbit, which is called the centripetal acceleration $(a_c)$.

$$a_c = \frac{v^2}{r} \quad (6)$$

Centripetal force ($f_c$) is a conceptual force and is defined by using mass of object (m) and centripetal acceleration ($a_c$), as in Newton's Second Law of motion.

$$F_c = ma_c \quad (7)$$

From equation (6) and (7), $F_c = \frac{mv^2}{r}$ (8)

$$F_c = \frac{mv^2}{r} = \frac{m(r\omega)^2}{r} = mr\omega^2 \quad (9)$$

$$F_c = mr(2\pi f)^2 = 4\pi mrf^2 \quad (10)$$

The percent difference, which measures precision of two experimental values, can be found using two experimental values of A and B.

$$PD = \frac{|A-B|}{(A+B)/2} \times 100\% \quad (11)$$

## APPARATUS AND PROCEDURE

- To perform this virtual experiment, you will need to access the simulation below: https://www.thephysicsaviary.com/Physics/Programs/Labs/ClassicCircularForceLab/

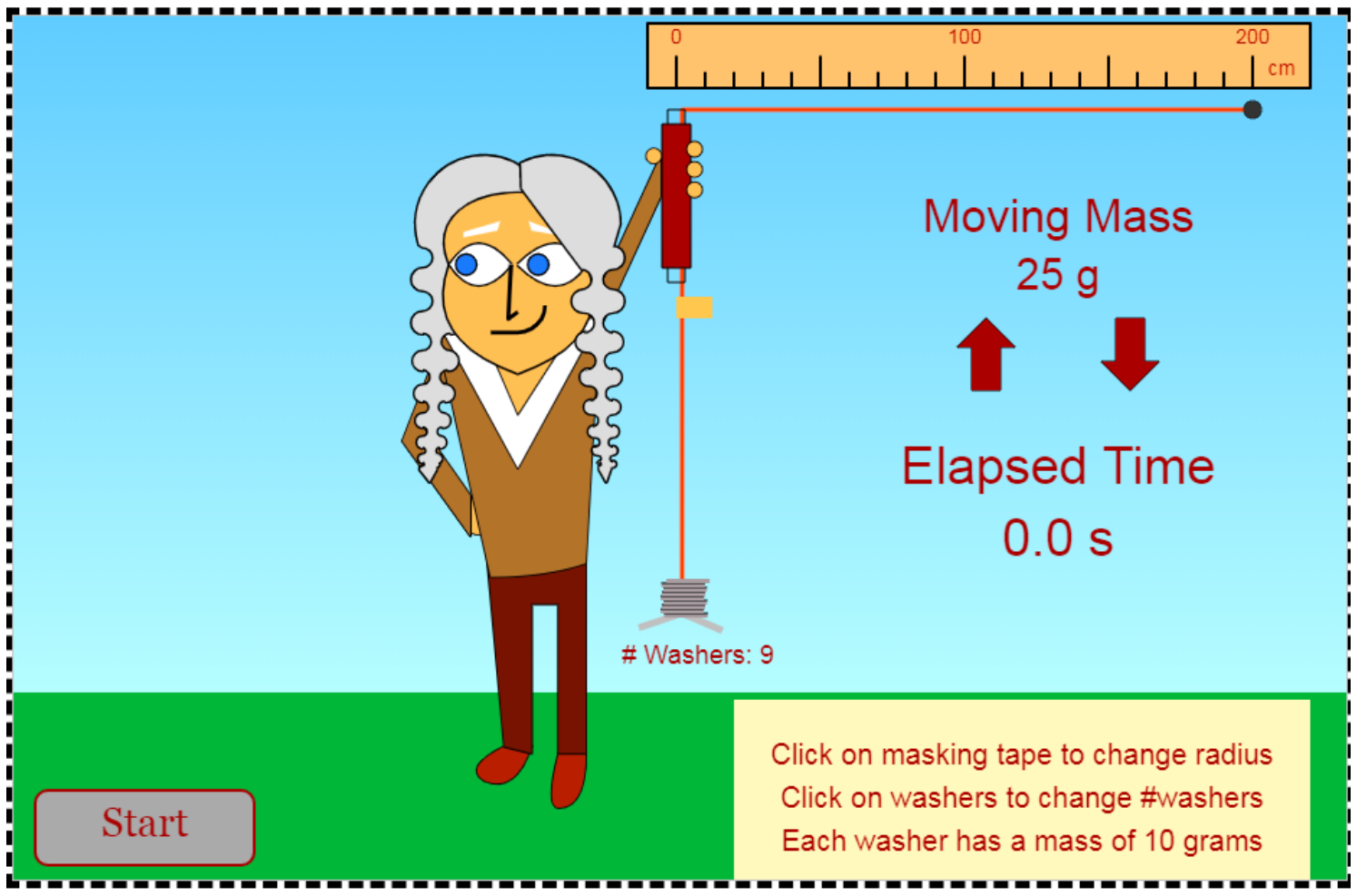

Figure 1 Centripetal force simulation (Picture credit, https://www.thephysicsaviary.com)

- A very detail video tutorial with data collection from simulator and data analysis with excel can be found here: https://youtu.be/WyrnRxO3cVE
- First draw a free-body-diagram (FBD) for the system. One FBD for each object.
- When you look at the free-body-diagram then you will see that the mass M in the horizontal string cannot stay like that unless the object is moving in a circle.
- When object M is moving in a horizontal circle then it can stay in the circle.
- Centripetal force on the mass M is provided by the tension of the string with which mass is attached.
- The same string goes through the tube and is attached to the hanger.
- Which means at the hanger, hanger weight equals the tension of the string.
- So, if the hanger is not moving, then the hanger weight should be equal to the centripetal force acting on the other end of the string.

### *Part I: The effect of mass of rotating object to centripetal force*

- First, we will check how the centripetal force changes as the mass M of the rotating object changes.
- Keep the radius R=2.0m and the mass of the hanger m=n*10.0*g* (n is number of washers in the hanger and each one is 10.0 grams) are fixed.
- Click start and count time for 5 revolutions. And repeat it three times.
- Increase mass M with the up arrow and click start. Count the time for 5 revolutions for three times.
- Record the information in table 1.
- First find the average time per one revolution.
- Calculate the angular velocity, tangential speed, centripetal acceleration, and centripetal force.
- Then compare the centripetal force from the experiment to the estimated value of it by using hanger mass.

### *Part II: The effect of radius of orbit of rotating object to centripetal force*

- Now will check the effect of the radius to centripetal force.
- Keep both the hanger mass (m) and rotating object mass (M) fixed.
- Click the yellow flag attached to the vertical string. It changes the radius of the orbit.
- Start this part of the experiment from about r=90.0cm.
- Click start and count time for 5 revolutions and repeat it three times.
- Record the data in table 2.
- Then by clicking the yellow flag change radius and repeat the last two steps above till the table 2 is complete.
- Then compare the centripetal force from the experiment to the estimated value of its by using hanger mass.

### *Part III: The effect of hanger mass to centripetal force*

- Now will keep both the radius of the circular path and the mass M of rotating object fixed but starts to change the hanger mass (m).
- When hanger mass changes it will change the centripetal force. As a result, the rotational speed of the object M changes.
- Start the hanger mass with about 10 washers in the hanger, radius of the path 2.0m and the mass M=25.0 $g$.
- Click on start and find the time for 5 revolutions. Repeat it 3 times.
- Record the data in table 3.
- Then increase the number of washers by clicking on the washer and repeat the last two steps above till the table 3 is complete.
- Calculate all the rotational quantities.
- Then complete the table 4.

## PRE-LAB QUESTIONS

1) What is the difference between frequency and angular velocity?
2) What is the correlation of period of oscillation and the centripetal acceleration?
3) What is the centripetal force?
4) Consider free-body-diagram of figure 1, which physical force creates the centripetal force of rotating object?
5) Consider the free-body-diagram of a figure 1, why is it very important to have the rotating string in level?

## DATA ANALYSIS AND CALCULATIONS

*Data tables, calculations and graphs must be done in excel. All the excel functions should be added below each table in the lab report.*

### *Part I: Effect of mass of the rotating object to centripetal force*

Hanger mass (m) = (# of washers) *10.0 grams

Radius of circular path (r) = 2.00 m

Centripetal force from hanger mass = $F_C$ =

Table 1 Rotational quantities as a function of mass M

| Mass, M [ ] | Time per 5 revolution [ ] | Time per one revolution T [ ] | $\omega = 2\pi f$ $\omega = \frac{2\pi}{T}$ [ ] | $v = r\omega$ [ ] | $a_c = \frac{v^2}{r}$ [ ] | Centripetal Force $F_{C_exp}$ [ ] | PD between $F_C$ and $F_{C_exp}$ |
|---|---|---|---|---|---|---|---|
| | | | | | | | |
| | | | | | | | |
| | | | | | | | |
| | | | | | | | |
| | | | | | | | |
| | | | | | | | |
| | | | | | | | |
| | | | | | | | |
| | | | | | | | |
| | | | | | | | |
| | | | | | | | |
| | | | | | | | |
| | | | | | | | |
| | | | | | | | |
| | | | | | | | |

- Discuss the effect of mass of the rotating object to rotational motion.

***Part II: Effect of radius of the circular path to centripetal force***

Hanger mass (m) = (# of washers)*10.0 grams

Mass of the object = M= 25.0*$10^{-3}$ kg

Centripetal force from hanger mass = $F_C$ =

Table 2 Rotational quantities as a function of radius r

<table>
<tr><th>Radius<br>r<br>[ ]</th><th>Time per 5<br>revolution<br>[ ]</th><th>Time per<br>one<br>revolution<br>T<br>[ ]</th><th>$\omega = 2\pi f$<br>$\omega = \frac{2\pi}{T}$<br>[ ]</th><th>$v = r\omega$<br>[ ]</th><th>$a_c = \frac{v^2}{r}$<br>[ ]</th><th>Centripetal<br>Force<br>$F_{C_exp}$<br>[ ]</th><th>PD<br>between<br>$F_C$<br>and<br>$F_{C_exp}$</th></tr>
<tr><td rowspan="3"></td><td></td><td rowspan="3"></td><td rowspan="3"></td><td rowspan="3"></td><td rowspan="3"></td><td rowspan="3"></td><td rowspan="3"></td></tr>
<tr><td></td></tr>
<tr><td></td></tr>
<tr><td rowspan="3"></td><td></td><td rowspan="3"></td><td rowspan="3"></td><td rowspan="3"></td><td rowspan="3"></td><td rowspan="3"></td><td rowspan="3"></td></tr>
<tr><td></td></tr>
<tr><td></td></tr>
<tr><td rowspan="3"></td><td></td><td rowspan="3"></td><td rowspan="3"></td><td rowspan="3"></td><td rowspan="3"></td><td rowspan="3"></td><td rowspan="3"></td></tr>
<tr><td></td></tr>
<tr><td></td></tr>
<tr><td rowspan="3"></td><td></td><td rowspan="3"></td><td rowspan="3"></td><td rowspan="3"></td><td rowspan="3"></td><td rowspan="3"></td><td rowspan="3"></td></tr>
<tr><td></td></tr>
<tr><td></td></tr>
<tr><td rowspan="3"></td><td></td><td rowspan="3"></td><td rowspan="3"></td><td rowspan="3"></td><td rowspan="3"></td><td rowspan="3"></td><td rowspan="3"></td></tr>
<tr><td></td></tr>
<tr><td></td></tr>
</table>

- Discuss the effect of radius of the circular path to rotational motion.

***Part III: Effect of hanger mass to rotational motion***

Mass of the object = M= 25.0*10$^{-3}$ kg

Radius = r = 2.0 m

Table 3 Rotational quantities as a function of hanger mass m

| # of washers on hanger | Hanger mass m [ ] | Time per 5 revolution [ ] | Time per one revolution T [ ] | $\omega = 2\pi f$ $\omega = \frac{2\pi}{T}$ [ ] | $v = r\omega$ [ ] | $a_c = \frac{v^2}{r}$ [ ] |
|---|---|---|---|---|---|---|
| | | | | | | |
| | | | | | | |
| | | | | | | |
| | | | | | | |
| | | | | | | |
| | | | | | | |
| | | | | | | |
| | | | | | | |
| | | | | | | |
| | | | | | | |
| | | | | | | |
| | | | | | | |
| | | | | | | |
| | | | | | | |
| | | | | | | |

Table 4 Centripetal force from hanger mass and from period of oscillation

| Hanger mass m [ ] | Centripetal acceleration $a_c$ [ ] | Centripetal force from hanger mass Fc(hanger) [ ] | Centripetal force from acceleration Fc($a_c$) [ ] | PD between Fc(hanger) and Fc($a_c$) |
|---|---|---|---|---|
| | | | | |
| | | | | |
| | | | | |
| | | | | |
| | | | | |

# EXPERIMENT 9 LINEAR MOMENTUM AND CONSERVATION LAWS

## OBJECTIVE

Linear momentum and conservation laws are investigated. Elastic and in elastic types of collisions are investigated. Theoretically calculated final velocities after collision are compared with experimentally calculated values.

## THEORY AND PHYSICAL PRINCIPLES

The linear momentum $\vec{p}$ of an object is represented by its mass (m) and velocity $\vec{v}$.

$$\vec{p} = m\vec{v} \tag{1}$$

Newton's second law of motion says that the force ($\vec{F}$) acting on an object is represented by its mass (m) and acceleration ($\vec{a}$).

$$\vec{F} = m\vec{a} \tag{2}$$

The force $\vec{F}$ can also be written as rate of change in momentum:

$$\vec{F} = \frac{d\vec{p}}{dt} \tag{3}$$

During the collision of a system of objects (net force acts on the system is zero), the change in momentum due to collision is zero for the system which means that the momentum is conserved.

$$\Delta\vec{p} = 0 \tag{4}$$

$$\Delta\vec{p} = \vec{p}_f - \vec{p}_i = 0 \tag{5}$$

$$\vec{p}_f = \vec{p}_i \tag{6}$$

Consider the collision of two objects, object 1 of mass $m_1$ and velocity $v_1$ and object 2 of mass $m_2$ and velocity $v_2$,

$$\vec{p}_{1i} + \vec{p}_{2i} = \vec{p}_{1f} + \vec{p}_{2f} \tag{7}$$

$$m_1\vec{v}_{1i} + m_2\vec{v}_{2i} = m_1\vec{v}_{1f} + m_2\vec{v}_{2f} \tag{8}$$

Collisions can be divided into two types as elastic collision in which energy is conserved and inelastic collision in which energy is not conserved.

An object in motion has kinetic energy (K), which is determined by its mass (m) and velocity (v).

$$KE = \frac{1}{2}mv^2 \tag{9}$$

Considering elastic cession and applying conservation of energy,

$$\Delta E(total) = 0 \tag{10}$$

$$\frac{1}{2}m_1v_{1f}^2 + \frac{1}{2}m_2v_{2f}^2 - \left(\frac{1}{2}m_1v_{1i}^2 + \frac{1}{2}m_2v_{2i}^2\right) = 0 \tag{11}$$

Consider elastic collision with object-1 initially moving and object-2 initially at rest, equation number 8 and 11 can be solved linearly to find the final velocities of objects after collision.

$$v_{1f} = \frac{(m_1 - m_2)}{(m_1 + m_2)} v_{1i} \qquad (12), \qquad\qquad v_{2f} = \frac{2m_1}{(m_1 + m_2)} v_{1i} \qquad (13)$$

In an inelastic collision where two objects stick to each other after collision, the conservation of momentum is rearranged to incorporate the one final velocity ($v_f$) for the combined masses.

$$m_1 \vec{v}_{1i} + m_2 \vec{v}_{2i} = (m_1 + m_2)\vec{v}_f \qquad (14)$$

The final velocity ($v_f$) after collision,

$$\vec{v}_f = \frac{m_1 \vec{v}_{1i} + m_2 \vec{v}_{2i}}{(m_1 + m_2)} \qquad (15)$$

In the laboratory experiment, velocities of the objects before and after collisions can be calculated by measuring time with a photogate. Then, conservation laws of momentum and energy can be investigated.

Velocity ($\vec{v}$) is calculated by the displacement ($\Delta\vec{x}$) and time ($\Delta t$).

$$\vec{v} = \frac{\Delta\vec{x}}{\Delta t} \qquad (16)$$

**APPARATUS AND PROCEDURE**

- This experiment is done with the following simulation. https://www.gigaphysics.com/momentum_lab.html
- A very detail video tutorial with data collection from simulator and data analysis with excel can be found here: https://youtu.be/Xdlx-XgEWZU

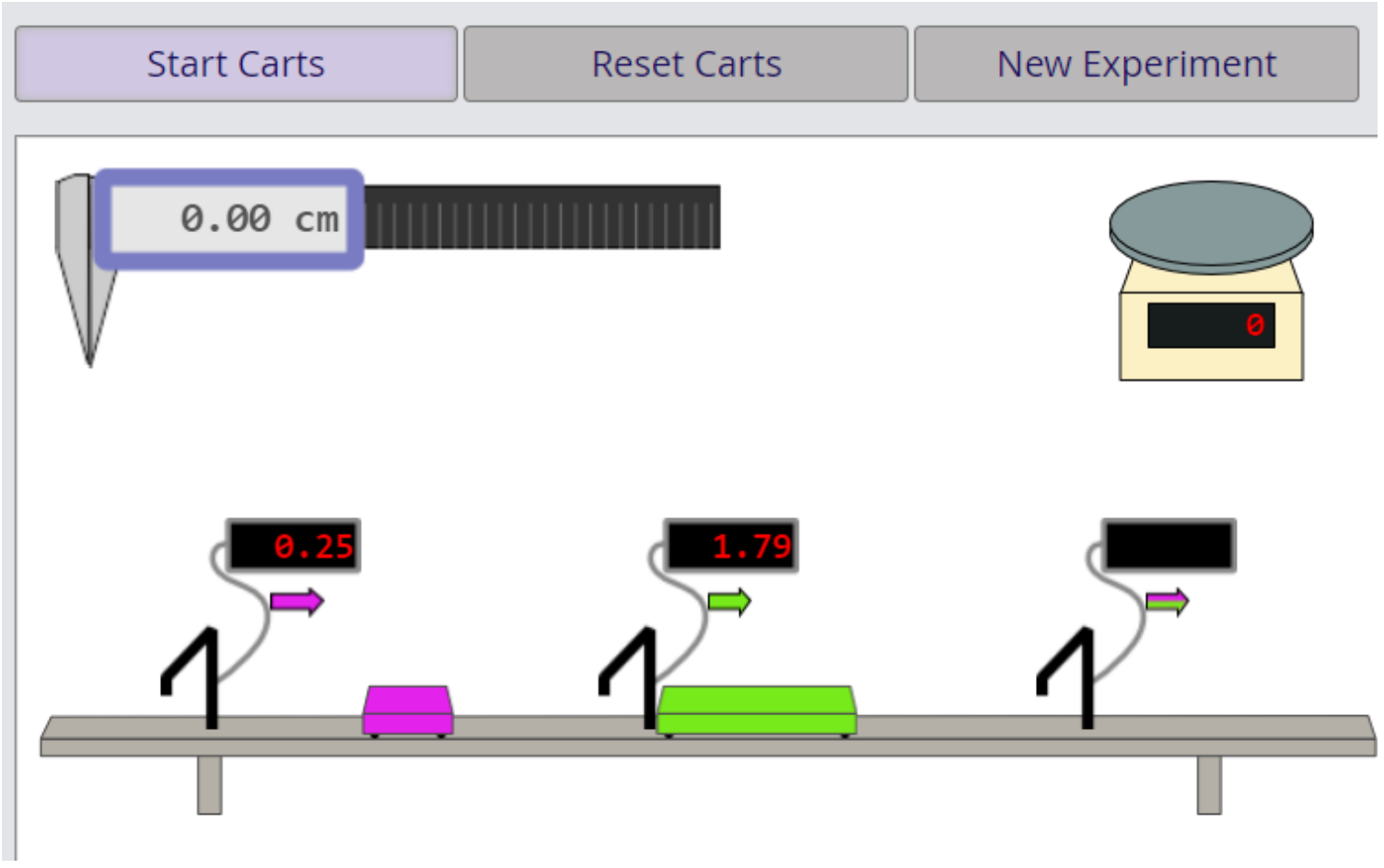

Figure 1 Simulator to study the momentum and conservation laws (Picture credit, https://www.gigaphysics.com)

***Part A: Elastic collision***

- Click and drag each cart and measure the mass of carts by placing them on the electronic balance on the top right corner.
- Click and drag each cart and measure the length of the carts by placing them on the vernier caliper on the top left corner.
- Set the type of collision to “elastic” by using the “cart behavior” menu.

- Set the “same direction” on the “cart’s direction” menu.
- Click on “start carts” and record the times on photogates with direction of motion. Consider the left to right is +x direction.
- Photogates have an arrow with color which indicates the cart number and the direction of motion.
- Set the “opposite direction” on the “cart’s direction” menu.
- Click on “start carts” and record the times on photogates with direction of motion. Consider the left to right is +x direction.
- Set the “one cart stationary” on the “cart’s direction” menu.
- Click on “start carts” and record the times on photogates with direction of motion. Consider the left to right is +x direction.

***Part B: Inelastic collision***

- Set the type of collision to “inelastic” by using the “cart behavior” menu.
- Set the “same direction” on the “cart’s direction” menu.
- Click on “start carts” and record the times on photogates with direction of motion. Consider the left to right is +x direction.
- Photogates have an arrow with color which indicates the cart number and the direction of motion.
- Set the “opposite direction” on the “cart’s direction” menu.
- Click on “start carts” and record the times on photogates with direction of motion. Consider the left to right is +x direction.
- Set the “one cart stationary” on the “cart’s direction” menu.
- Click on “start carts” and record the times on photogates with direction of motion. Consider the left to right is +x direction.

***Part C: Partially elastic collision***

- Set the type of collision to “partially elastic” by using the “cart behavior” menu.
- Set the “same direction” on the “cart’s direction” menu.
- Click on “start carts” and record the times on photogates with direction of motion. Consider the left to right is +x direction.
- Photogates have an arrow with color which indicates the cart number and the direction of motion.
- Set the “opposite direction” on the “cart’s direction” menu.
- Click on “start carts” and record the times on photogates with direction of motion. Consider the left to right is +x direction.
- Set the “one cart stationary” on the “cart’s direction” menu.
- Click on “start carts” and record the times on photogates with direction of motion. Consider the left to right is +x direction.

**PRE LAB QUESTIONS**

1) What is meant by the system in the case of conservation of momentum?
2) If a system contains two objects, is there a possibility that total momentum of one object is not conserved? Explain your answer?
3) Explain how to find the impulse that acts on one object during the collision?
4) What are the types of collisions?
5) Describe the types of collisions by suing conservation laws?

**DATA ANALYSIS AND CALCULATIONS**

*Data tables, calculations and graphs must be done in excel. All the excel functions should be added below each table in the lab report.*

Table 1 Time measurements of carts before and after collision

| Cart 1 | | Cart 2 | |
|---|---|---|---|
| Mass, m1 [ ] | Length, L1 [ ] | Mass, m1 [ ] | Length, L1 [ ] |
| | | | |

| | Time before collision | | Time after collision | |
|---|---|---|---|---|
| | Cart 1 | Cart 2 | Cart 1 | Cart 2 |
| | $t_{1i}$ [ ] | $t_{2i}$ [ ] | $t_{1f}$ [ ] | $t_{2f}$ [ ] |
| Elastic collision Carts moves to same direction | | | | |
| Elastic collision Carts moves to Opposite direction | | | | |
| Elastic collision One cart at rest | | | | |
| Inelastic collision Cart moves to same direction | | | | |
| Inelastic collision Cart moves to opposite direction | | | | |
| Inelastic collision One cart at rest | | | | |
| Partially elastic collision Cart moves to same direction | | | | |
| Partially elastic collision Cart moves to opposite direction | | | | |
| Partially elastic collision One cart at rest | | | | |

Table 2 Analysis of final velocities with theoretical and experimental

| | Velocity before collision | | Velocity after collision | |
|---|---|---|---|---|
| | Cart 1 | Cart 2 | Cart 1 | Cart 2 |
| | $v_{1i}$ [ ] | $v_{2i}$ [ ] | $v_{1f}$ [ ] | $v_{2f}$ [ ] |
| Elastic collision Carts moves to same direction | | | | |
| Elastic collision Carts moves to Opposite direction | | | | |
| Elastic collision One cart at rest | | | | |
| Inelastic collision Cart moves to same direction | | | | |
| Inelastic collision Cart moves to opposite direction | | | | |
| Inelastic collision One cart at rest | | | | |
| Partially elastic collision Cart moves to same direction | | | | |
| Partially elastic collision Cart moves to opposite direction | | | | |
| Partially elastic collision One cart at rest | | | | |

Table 3 Analysis of momentum conservation

| | $P_i$ *(total)* Initial Total [ ] | $P_f$ *(total)* Final Total [ ] | Changing Momentum ΔP [ ] | Does the momentum conserved or not? Explain why or why not? |
|---|---|---|---|---|
| Elastic collision Carts moves to same direction | | | | |
| Elastic collision Carts moves to Opposite direction | | | | |
| Elastic collision One cart at rest | | | | |
| Inelastic collision Cart moves to same direction | | | | |
| Inelastic collision Cart moves to opposite direction | | | | |
| Inelastic collision One cart at rest | | | | |
| Partially elastic collision Cart moves to same direction | | | | |
| Partially elastic collision Cart moves to opposite direction | | | | |
| Partially elastic collision One cart at rest | | | | |

Table 4 Analysis of energy conservation

| | $E_i$ *(total)* Initial Total [ ] | $E_f$ *(total)* Final Total [ ] | Changing Energy ΔE [ ] | Does the energy conserve or not? Explain why or why not? |
|---|---|---|---|---|
| Elastic collision Carts moves to same direction | | | | |
| Elastic collision Carts moves to Opposite direction | | | | |
| Elastic collision One cart at rest | | | | |
| Inelastic collision Cart moves to same direction | | | | |
| Inelastic collision Cart moves to opposite direction | | | | |
| Inelastic collision One cart at rest | | | | |
| Partially elastic collision Cart moves to same direction | | | | |
| Partially elastic collision Cart moves to opposite direction | | | | |
| Partially elastic collision One cart at rest | | | | |

# EXPERIMENT 10

# TORQUE AND STATIC EQUILIBRIUM

## OBJECTIVE

An object at static equilibrium will be studied in detail. Newton laws and net torque at a given point will be applied to a large object at static equilibrium. Center of mass and the object and its importance to static equilibrium of large objects will be investigated.

## THEORY AND PHYSICAL PRINCIPLES

If a rigid body is in equilibrium, then the vector sum of the external forces acting on the body yields a zero resultant and the sum of the torques of the external forces about any arbitrary axis is also equal to zero.

Stated in equation form: $\Sigma F = 0$ and $\Sigma\tau = 0$ (1)

In this lab force of gravity also will be needed:

$$F = mg \qquad (2)$$

where m is in kg and $g$ in m/s$^2$ therefore, force F is in Newton N.

Torque is a measure of how effective a given force is at twisting or turning the object it is applied to. Torque is defined as the force F times the moment arm or lever arm r of the force with respect to a selected pivot point x. In other words, r is the distance from the pivot to the point where a force is applied.

$\hat{n}$ is the unit vector and $\theta$ is the angle between lever arm and the force.

$$\sum \vec{\tau}_a = \sum \vec{r}_a \otimes \vec{F}_a$$

$$\vec{\tau} = r\ F\ sin\theta\ \hat{n} \qquad (3)$$

When the lever arm is perpendicular to the force,

$$\vec{\tau} = r\ F\ \hat{n} \qquad (4)$$

Usually lever arm and the force are on xy-plane (on the paper) and the resultant torque is on z-axis, either coming out the paper (+z direction, with $+\tau$, counter-clock-wise rotation) or into the paper (-z direction, with $-\tau$, clock-wise rotation).

So, in this experiment magnitude of the resultant torque due to each force will be represented by,

$$\tau = r\ F \qquad (5)$$

The unit of torque is meter-Newton, mN.

In this experiment a meter stick is used as a rigid body to illustrate the application of the equations of equilibrium. For this experiment it is not only important to familiarize yourself with the equations, but also with sketching free-body diagrams (FBD).

## *APPARATUS AND PROCEDURE*

- Following items are for in-class experiments.
- Meter stick
- One knife-edge meter stick clamp without clips
- Two knife-edge meter stick clamps with clips
- Two 50 gram hangers
- Slotted weights
- Meter stick support stand
- Balance

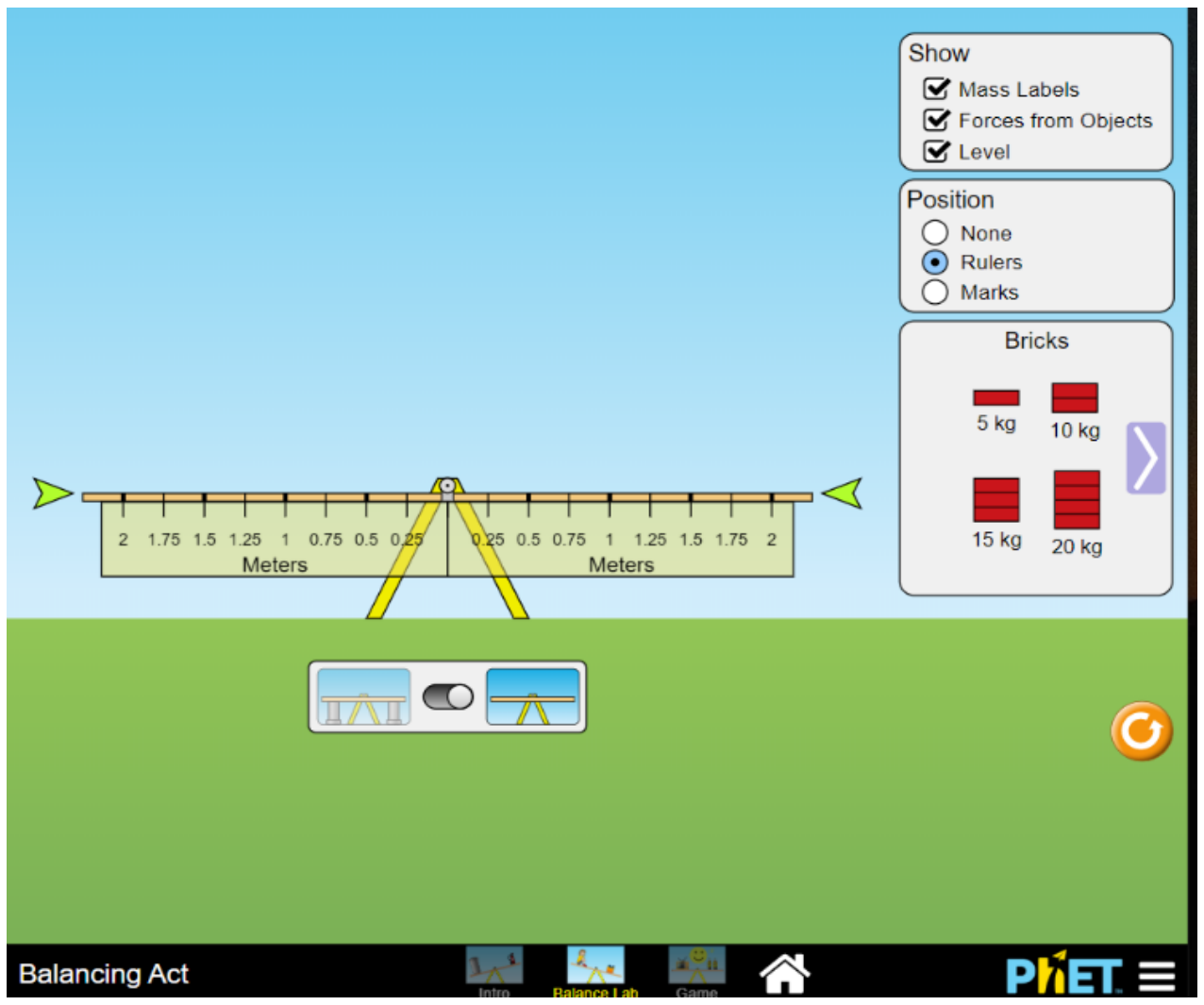

Figure 1  Simulator for equilibrium experiment (Picture credit, https://phet.colorado.edu)

- This experiment is done with following simulation, https://phet.colorado.edu/en/simulation/balancing-act
- A very detail video tutorial with data collection from simulator and data analysis with excel can be found here: https://youtu.be/WqyQeuGJ94U

### *Part A: Meter stick (Meter stick in the simulator is 2.00m long) in equilibrium, fulcrum at its center and masses on opposite sides of the balance point.*

- Fulcrum (pivot point) should be exactly at the center of mass point of the ruler.
- Add a mass of $m_1$=5.0kg at x=1.00m location on positive x direction.
- Meter-stick can be balanced by using a second mass adding to the other side of the meter stick. Calculate the position of the mass $m_2$=10.0kg by using the torque balance equation.
- Then balance the ruler by adding a mass of $m_2$=10.0 kg to the other side (negative x direction) of the pivot point.
- Compare the experimental balance point $x_2$ of the mass $m_2$ with calculated $x_2$?
- Calculate clock-wise (cw) torque at pivot point $\sum \tau_{cw}$, this is a negative value.
- Calculate counter-clock-wise (ccw) torque at pivot point $\sum \tau_{ccw}$, this is a positive value.
- Then check the equilibrium condition by using net torque at pivot point.
- Then repeat above measurements for $m_1$=5kg at x=1.00m and $m_2$=20.0kg.

### *Part B: Balancing the meter stick with multiple forces*

- Add a person with mass of $m_1$=20.0kg at $x_1$=1.00m location on negative x direction and another person with mass of $m_2$=80.0kg at $x_2$=0.75m location on positive x direction.
- Meter-stick can be balanced by adding third mass $m_3$=20.0kg to the same side as the mass $m_1$. Calculate the position of the mass $m_3$ by using the torque balance equation.
- Then balance the ruler by adding a mass of $m_3$ experimentally.

- Calculate clock-wise (cw) torque at pivot point $\sum \tau_{cw}$, this is a negative value.
- Calculate counter-clock-wise (ccw) torque at pivot point $\sum \tau_{ccw}$, this is a positive value.
- Repeat the above procedure by changing the second person with 60.0kg.

***Part C: Find the unknown mass by using the method of torque.***

- Add a mass of $m_1$=unknown-mass-A at x=1.00m location on positive x direction.
- Then balance the ruler by adding a mass of $m_2$=20.0 kg to the other side (negative x direction) of the pivot point.
- And find the exact balance point $x_2$ of the mass $m_2$.
- Then use the net torque equation to find the unknown mass of the object.
- Then repeat the above measurements for unknown object B, C and D.

***Part D: Find the mass of the meter ruler by using torque and graphical method***

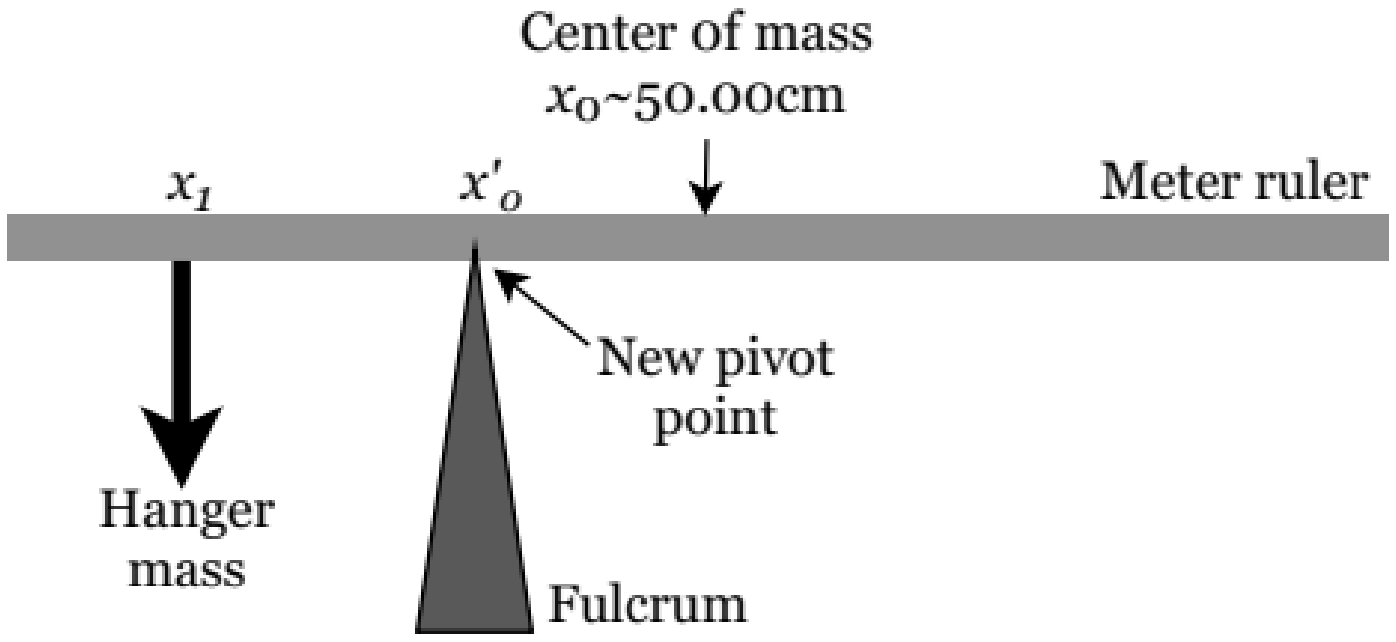

Figure 2 Experiment setup to investigate the mass of the meter stick

- This part of the experiment cannot be done with simulation. Therefore, in class real data will be used to do analysis.
- In class data sets can be found in table 4.
- Following steps are for in class experiments only.
- Pivot point should be 30.0cm location.
- Add a hanger of mass $m_1$=50.0 gram at the left side (towards the zero side of the ruler) of the new pivot point.
- Then balance the meter ruler by adjusting the position of the hanger of mass $m_1$.
- Note down the mass m1 and position of hanger $x_1$ on the data table.
- Then, add 20.0 gram to the hanger, and change the position of the hanger till the meter ruler is balanced.
- Note down the mass $m_1$ and position of the hanger $x_1$ on the data table.
- Repeat the last two steps above by adding 20.0 gram at a time for four more steps.
- Make a graph of $|x_0' - x_1|\ vs\ \frac{1}{m_1}$ .
- Calculate the mass of the meter ruler (M) by using the slope of the graph.

**PRE LAB QUESTIONS**

1) Describe center of mass of an object?
2) Describe the torque?
3) Describe static equilibrium?
4) Consider figure 2 and describe all the forces acts on the meter ruler?
5) Consider figure 2 and describe torque balance equation?

**DATA ANALYSIS AND CALCULATIONS**

*Data tables, calculations and graphs must be done in excel. All the excel functions should be added below each table in the lab report.*

**Part A:** ***Meter stick in equilibrium, fulcrum at its center and masses on opposite sides of the balance point***

Table 1 Meter stick at equilibrium with two masses

<table>
<tr><th colspan="2"></th><th>Case 1</th><th>Case 2</th></tr>
<tr><td>Mass (m2) on right side</td><td>[ ]</td><td></td><td></td></tr>
<tr><td>Position (x1) of mass m1</td><td>[ ]</td><td></td><td></td></tr>
<tr><td>Mass (m2) on left side</td><td>[ ]</td><td></td><td></td></tr>
<tr><td>Position (x2) of mass m2 calculated</td><td>[ ]</td><td></td><td></td></tr>
<tr><td>Position (x2) of mass m2 experiment</td><td>[ ]</td><td></td><td></td></tr>
<tr><td>Counter clock-wise torque, τ(ccw)</td><td>[ ]</td><td></td><td></td></tr>
<tr><td>Clock-wise torque, τ(cw)</td><td>[ ]</td><td></td><td></td></tr>
<tr><td>Net torque τ(net)</td><td>[ ]</td><td></td><td></td></tr>
<tr><td colspan="4">Picture of torque balance case 1</td></tr>
<tr><td colspan="4"></td></tr>
<tr><td colspan="4">Picture of torque balance case 2</td></tr>
<tr><td colspan="4"></td></tr>
</table>

- Discuss the net torque equation of the equilibrium object?

**Part B:** ***Meter stick in equilibrium with three masses***

Table 2 Meter stick at equilibrium with three masses

| | | Case 1 | Case 2 |
|---|---|---|---|
| Mass (m1) of person on right side | [ ] | | |
| Position (x1) of mass m1 | [ ] | | |
| Mass (m2) of person on right side | [ ] | | |
| Position (x2) of mass m2 | [ ] | | |
| Counter clock-wise torque, τ(ccw) | [ ] | | |
| Clock-wise torque, τ(cw) | [ ] | | |
| Torque needed to balance τ(needed) | [ ] | | |
| Position of the m3=20.0kg mass needed to balance | x3(cal) | | |
| | x3(exp) | | |
| Picture of torque balance case 1 | | | |
| | | | |
| Picture of torque balance case 2 | | | |
| | | | |

- Discuss the net torque equation of the equilibrium object?

**Part B:** ***Finding unknown mass by using the method of torque***

Table 3 Unknown by using method of torques

| | | Case A | Case B | Case C | Case D |
|---|---|---|---|---|---|
| Position of unknown mass x(u) | [ ] | | | | |
| Mass on left side m | [ ] | | | | |
| Position of mass m, x | [ ] | | | | |
| Counter clock-wise torque, τ(ccw) | [ ] | | | | |
| Clock-wise torque, τ(cw) | [ ] | | | | |
| Unknown mass by using method of torque, m(u) | [ ] | | | | |

| Picture of torque balance case A |
|---|
| |
| Picture of torque balance case B |
| |
| Picture of torque balance case C |
| |
| Picture of torque balance case D |
| |

- Discuss the application of net torque for this part of the experiment?

**Part D: *Obtaining the mass of the meter stick by method of torques***

Table 4 Measurements to find the mass of the meter stick by graphical method.

| Total mass in hanger-1 $m_1$ [ kg ] | Position of the hanger-1 $x_1$ [ m ] | $\|x_0 - x_1\|$ [ ] | Mass of the meter stick $M_{cal}$ [ ] | $\frac{1}{m_1}$ [ ] | Percent difference between $M_{cal}$ vs M |
|---|---|---|---|---|---|
| 0.06641 | 0.0850 | | | | |
| 0.08641 | 0.1345 | | | | |
| 0.10641 | 0.1650 | | | | |
| 0.12641 | 0.1865 | | | | |
| 0.14641 | 0.2020 | | | | |
| Draw a free body diagram (FBD) for balancing the meter ruler at new pivot point. Add a picture of FBD here. | | | | | |

- Find the average of the calculated mass of the meter stick, $M_{cal_avg}$ =
- Make a graph of $|x_0' - x_1|\ vs\ \frac{1}{m_1}$ . Calculate the mass of the meter ruler (M) by using the slope of the graph.

  M_graph =

- Find the percent difference (PD) between following:

  PD (M_cal_avg vs M_graph) =

# EXPERIMENT 11 HOOKE'S LAW AND SIMPLE HARMONIC MOTION

## OBJECTIVE

In this experiment, you will investigate Hooke's Law as well as the parameters within the concept of simple harmonic motion. Upon completion of this experiment, you will be able to graphically and analytically interpret Hooke's Law and will understand the concept of simple harmonic motion.

## THEORY AND PHYSICAL PRINCIPLES

Hooke's law: $F = k\Delta y$ (1)

F is applied force and $\Delta y$ is the extension or compression of the spring. These two quantities are opposite directions to each other. There is an internal force in spring which is called spring force which is opposite to the applied force.

Spring constant: $k = \frac{F}{\Delta y}$ (2)

Ratio between spring force and the extension/compression of the spring is constant up to a maximum applied force limit. This limit is called the elastic limit in which when the applied force is removed spring comes back to its original (relaxed) length.

Simple Harmonic Motion (SHM)

When an object moves in a straight but back and forth around a fixed point, this motion is called a SHM. Energy is conserved in this motion and if energy starts to lose then the SHM becomes damped harmonic oscillations.

Simple harmonic motion of a spring:

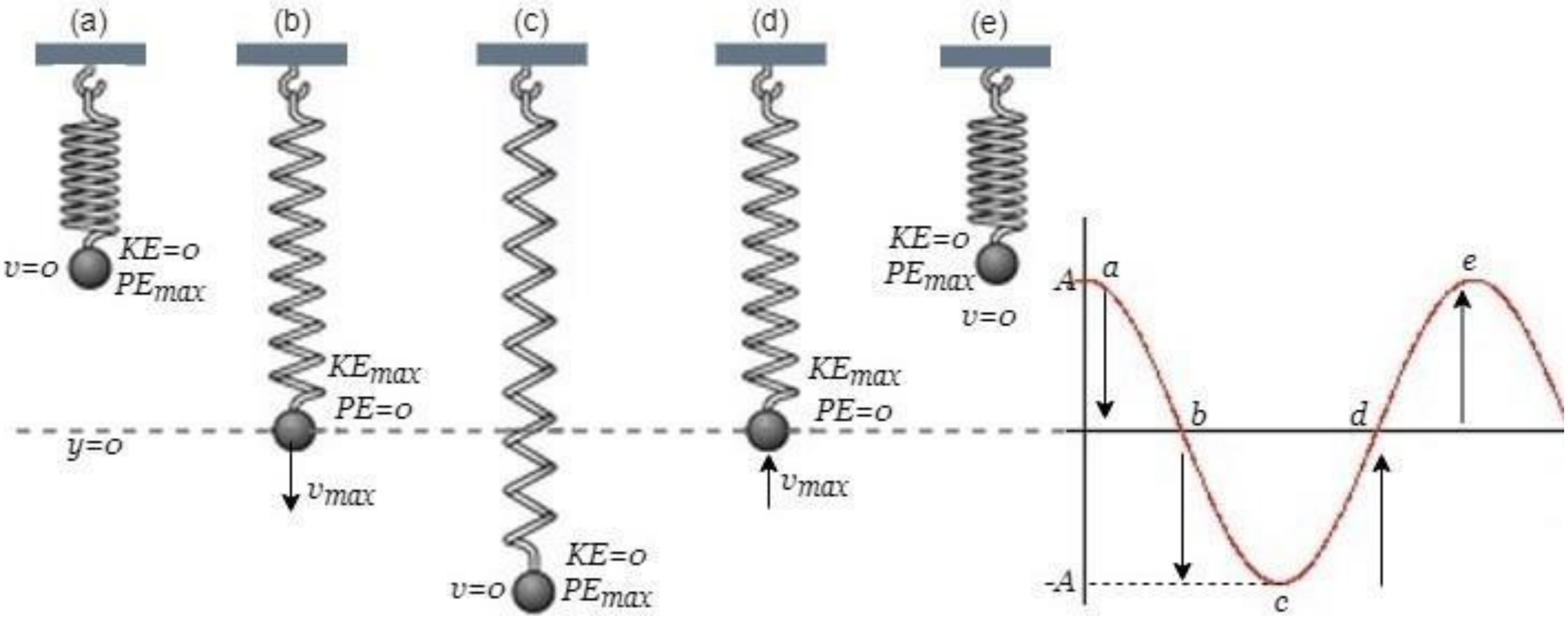

Figure 1 SHM of a spring and path of motion as a cosine function

Time for the one complete oscillation of the spring is called the period of oscillation (T) and how many oscillations per second is called the frequency (f) of the motion.

The frequency f and period T are related by: $f = \frac{1}{T}$ (3)

By applying Hooke's law and Newton's second law of motion at any given moment of motion of the object,

$$F = ma = -kx \quad (4)$$

$$a = -\frac{k}{m}x \quad (5)$$

Equation of simple harmonic motion,

$$a = -\omega^2 x \quad (6)$$

From equations (5) and (6),

$$\omega = \sqrt{\frac{k}{m}} \quad (7)$$

$$\omega = 2\pi f = \frac{2\pi}{T} \quad (8)$$

For a simple spring mass system, the dependence of the frequency and period on the spring constant, k and the mass, m is given by:

$$f = \frac{1}{2\pi}\sqrt{\frac{k}{m}} \quad \text{or} \quad T = 2\pi\sqrt{\frac{m}{k}} \quad (9)$$

$$T^2 = \left(\frac{4\pi^2}{k}\right)m \quad (10)$$

## APPARATUS AND PROCEDURE

- This experiment will be done completely with the following simulation. Please click here to access the simulation. https://phet.colorado.edu/sims/html/masses-and-springs/latest/masses-and-springs_en.html
- A very detail video tutorial with data collection from simulator and data analysis with excel can be found here: https://youtu.be/UDU90VE6Gho

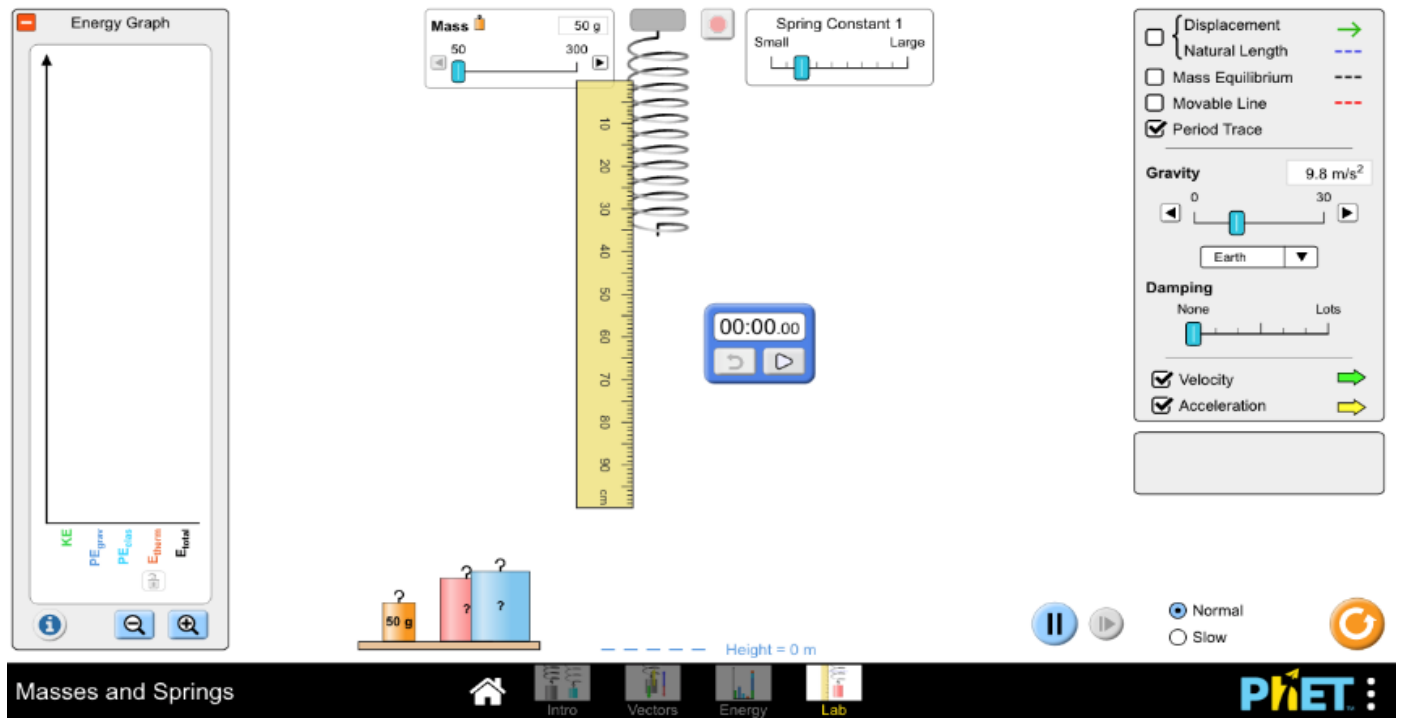

Figure 2 Simulation for Hooke's law and simple-harmonic-motion (Picture credit, https://phet.colorado.edu)

***Hooke's law and simple-harmonic-oscillation (SHM) of spring***

- Set up the spring-1 by using the spring constant into small (left side of spring constant scale) as shown in the simulation.
- Make sure the damping scale in the simulation is in the position of "none".
- Set the mass to $50g$ level of the mass changing scale of the simulation.
- Place ruler very close to spring and the hanger mass and note down the very bottom position ($y_0$) of the hanger.
- Add the mass $50g$ weight to the spring.
- Record the new position $y_1$ in the table 1.
- Then bring the timer very close to mass attached to spring.
- Give small oscillation by dragging the mass attached to spring about 1.0cm and release it.
- Switch on the timer and count the time for 5 oscillations and record it in the table 2.
- Since the oscillations are faster it is difficult to count the number of oscillations. Therefore, use the "slow motion" in the simulator. It will not affect the time and timer gives the correct time.
- Repeat all the above procedures for another six masses with increasing $50g$ at a time. Use the mass scalar to increase the mass attached to spring.
- Make a graph of applied force (m$g$) vs $|y_1 - y_0|$
- Find the spring constant by using the slope of the graph.
- Make a graph of applied mass (m) vs period of oscillation (T).
- Find the spring constant by using the slope of the graph.
- Repeat all the above procedures for large spring by changing the spring constant scale into far right as shown in the simulation.

**PRE LAB QUESTIONS**

1) Describe spring force?
2) Describe spring constant?
3) Describe spring potential energy?
4) Consider oscillating spring with mass attached, if the mass of the object doubled, what will happen to period of the oscillation?
5) Consider oscillating spring, what is the effect of the gravitational constant to the period of the oscillation of the spring?

## DATA ANALYSIS AND CALCULATIONS

*Data tables, calculations and graphs must be done in excel. All the excel functions should be added below each table in the lab report.*

***Hooke's law of spring-1*** *(small spring, change the scale of spring constant to far left side on simulation)*

Table 1 Measurements for the spring constant of spring-1

| Mass (m) [ ] | Force (F=m$g$) [ ] | Y-scale reading ($y_1$) [ ] | $\Delta y = \lvert y_i - y_0 \rvert$ [ ] | $k = \frac{F}{\Delta y}$ [ ] |
|---|---|---|---|---|
| | | | | |
| | | | | |
| | | | | |
| | | | | |
| | | | | |
| | | | | |

- Calculate the average value of spring constant ($k_{avg}$) by averaging the last column of above table?
  $k_{avg}$ =
- Find the spring constant ($k_{graph1}$) by using the slope of the graph of force (mg) vs $\Delta y$?
  $k_{graph1}$ =
- Calculate the percent difference between $k_{avg}$ and $k_{graph1}$ ?

***SHM behavior of spring-1*** *(Small spring, change the scale of spring constant to far left side on simulation)*

Note: Time in the stopwatch in simulator is written in, t = 00:00.00 = (hrs)(min):(sec)(sec/10)

Use the "slow motion" in the simulator. It will not affect the time and timer gives the correct time.

Table 2 Measurements for the spring constant by using SHM

| Mass (M) [ ] | Period for 5 Oscillations [ ] | Time for one Osci. $T_{avg}$ [ ] | $T^2_{avg}$ [ ] | $k_{SHM} = \frac{4\pi^2 m}{T^2}$ [ ] |
|---|---|---|---|---|
| | | | | |
| | | | | |
| | | | | |
| | | | | |
| | | | | |
| | | | | |
| | | | | |
| | | | | |
| | | | | |
| | | | | |
| | | | | |
| | | | | |
| | | | | |
| | | | | |
| | | | | |
| | | | | |
| | | | | |
| | | | | |

- Calculate the average value of spring constant ($k_{avg}$) by averaging the last column of above table?
  $k_{avg}$ (SHM) =
- Make graph of $T^2$ vs $m$. Calculate the spring constant $k_{graph}$(SHM) by using the slope of the graph?
  $k_{graph}$(SHM) =
- Calculate the percent difference between $k_{avg}$ and $k_{avg}$(SHM)?
- Calculate the percent difference between $k_{graph}$ and $k_{graph}$(SHM)?

***Hooke's law of spring-2*** *(Large spring, change the scale of spring constant to far right side on simulation)*

Table 3 Measurements for the spring constant of spring-2

| Mass (m) [ ] | Force (F=m$g$) [ ] | Y-scale reading ($y_1$) [ ] | $\Delta y = \lvert y_i - y_0 \rvert$ [ ] | $k = \frac{F}{\Delta y}$ [ ] |
|---|---|---|---|---|
| | | | | |
| | | | | |
| | | | | |
| | | | | |
| | | | | |
| | | | | |

- Calculate the average value of spring constant ($k_{avg}$) by averaging the last column of above table?
  $k_{avg}$ =
- Find the spring constant ($k_{graph1}$) by using the slope of the graph of force (mg) vs $\Delta y$?
  $k_{graph1}$ =
- Calculate the percent difference between $k_{avg}$ and $k_{graph1}$ ?

***SHM behavior of spring-2*** *(Small spring, change the scale of spring constant to far right side on simulation)*

Note: Time in the stopwatch in simulator is written in, t = 00:00.00 = (hrs)(min):(sec)(sec/10)

Use the "slow motion" in the simulator. It will not affect the time and timer gives the correct time.

Table 4 Measurements for the spring constant by using SHM

| Mass (m) [ ] | Period for 5 Oscillations [ ] | Time for one Osci. $T_{avg}$ [ ] | $T_{avg}^2$ [ ] | $k_{SHM} = \frac{4\pi^2 m}{T^2}$ [ ] |
|---|---|---|---|---|
| | | | | |
| | | | | |
| | | | | |
| | | | | |
| | | | | |
| | | | | |
| | | | | |
| | | | | |
| | | | | |
| | | | | |
| | | | | |
| | | | | |
| | | | | |
| | | | | |
| | | | | |
| | | | | |
| | | | | |
| | | | | |

- Calculate the average value of spring constant ($k_{avg}$) by averaging the last column of above table?
  $k_{avg}$ (SHM) =
- Make graph of $T^2$ vs $m$. Calculate the spring constant $k_{graph}$(SHM) by using the slope of the graph?
  $k_{graph}$(SHM) =
- Calculate the percent difference between $k_{avg}$ and $k_{avg}$(SHM)?
- Calculate the percent difference between $k_{graph}$ and $k_{graph}$(SHM)?

# EXPERIMENT 12
# SIMPLE PENDULUM

## OBJECTIVE

- Apply the scientific method to theoretical prediction to check their validity.
- Understand how physical parameters are varied so as to investigate theoretical predictions.
- Appreciate the use of approximation to facilitate investigation and analysis.
- Use the period of a simple pendulum to investigate the gravitational acceleration.

## THEORY AND PHYSICAL PRINCIPLES

Period of the simple pendulum is given by following equation and T is period (time for one complete oscillation), L is length of the string (measured from top-contact point to the middle of the mass attached) and $q$ is the angular displacement.

$$T = 2\pi\sqrt{\frac{L}{g}}\left(1+\frac{1}{4}sin^2\frac{\theta}{2}+\frac{9}{64}sin^4\frac{\theta}{2}+\cdots\dots\dots\dots..\right) \quad (1)$$

Will use the approximation with only first two terms as follows:

$$T_{cal1} = 2\pi\sqrt{\frac{L}{g}}\left(1+\frac{1}{4}sin^2\frac{\theta}{2}\right) \quad (2)$$

When the angular displacement is small (about 20 degrees or less), about series expansion can be reduced into following:

$$T_{cal2} = 2\pi\sqrt{\frac{L}{g}} \quad (3)$$

Period of the simple pendulum can be predicted by using above theoretical formulas. Also, by measuring period and length of the pendulum, one can calculate gravitational acceleration in the location by manipulating above equation (3),

$$g = \frac{4\pi^2 L}{T^2} \quad (4)$$

Gravitational acceleration also can be found by using a graphical method.

$$T = 2\pi\sqrt{\frac{L}{g}} \quad (5)$$

$$T^2 = \frac{4\pi^2}{g}L \quad (6)$$

By using the slope of the graph of $T^2$ vs L,

$$g = \frac{4\pi^2}{slope} \quad (7)$$

Period average, $$\bar{T}_{exp} = \frac{\sum_I^n T_i}{N} \quad (8)$$

Mean deviation of average period of oscillation, $\delta\bar{T} = \frac{\sum_i^N |T-\bar{T}|}{N}$ (9)

## APPARATUS AND PROCEDURE

- This experiment is done with the following simulation: https://phet.colorado.edu/en/simulation/pendulum-lab

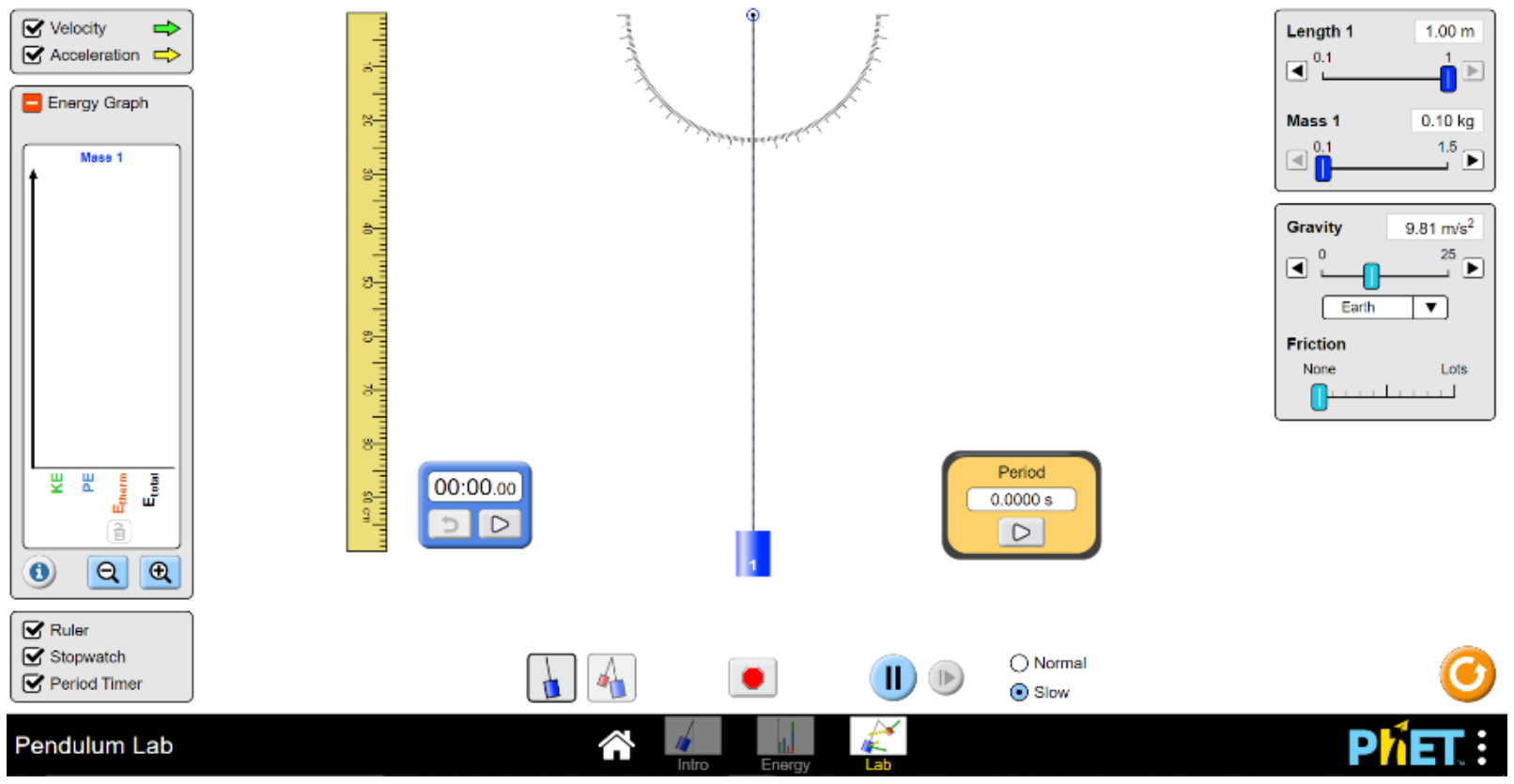

Figure 1 Simple pendulum simulation (Picture credit, https://phet.colorado.edu)

### *Part A: Investigation of small angle approximation*

- Measure time for 5 periods of oscillations by using a stopwatch as a function of angle.
- Calculate period by using theoretical equations (1) and (2).
- Compare the results by calculating percent error.

### *Part B: Investigation of Period of Oscillation with changing Mass*

- Use $10^0$ angle and fixed length (L) for this part of the experiment.
- Measure time for 5 periods of oscillations by using stopwatch as a function mass.
- Compare the results by calculating percent error.

### *Part C: Investigation of Length effect and calculate gravitational constant*

- Use $10^0$ angle and fixed mass (M) for this part of the experiment.
- Measure time for 5 periods of oscillations by using stopwatch as a function length.
- Make a graph of T vs L and fit with a 2nd order polynomial.
- Explain the behavior of this graph in the report.
- Make a graph of $T^2$ vs L and do the linear fitting.
- Explain the behavior of this graph in the report.
- Calculate experimental gravitation constant by using equation-4 (use measured values of T and L)
- Calculate average value for experimentally calculated gravitational constant, $g_{avg_exp}$.
- Find the gravitational constant ($g_{slope}$) by using the slope of the graph of $T^2$ vs L?

## PRE LAB QUESTIONS

1) Period of the simple pendulum depends on what physical quantities?
2) What is the effect of mass attached to the period of the pendulum?
3) If you shorten the length of the pendulum what will happen to the period?
4) Give the locations of the motion of the pendulum with respect to following conditions,
   a) Maximum displacement?
   b) Maximum velocity?
   c) Maximum acceleration?

**DATA TABLES AND ANALYSIS**

*Data tables, calculations and graphs must be done in excel. All the excel functions should be added below each table in the lab report.*

***Part A: Investigation of small angle approximation***

Mass = Length =

Table 1 Analysis of period of oscillation with changing angle

| Angle $q$ [ ] | Time for 5 Osci. [ ] | Period $T_{exp}$ [ ] | Period $T_{cal-1}$ [ ] | Period $T_{cal-2}$ [ ] | PD $T_{cal-1}$ & $T_{cal-2}$ [ ] | PD $T_{exp}$ & $T_{cal-1}$ [ ] | PD $T_{exp}$ & $T_{cal-2}$ [ ] |
|---|---|---|---|---|---|---|---|
| $5^0$ | | | | | | | |
| $10^0$ | | | | | | | |
| $20^0$ | | | | | | | |
| $30^0$ | | | | | | | |
| $40^0$ | | | | | | | |
| $50^0$ | | | | | | | |
| $60^0$ | | | | | | | |

***Part B: Investigation of Period of Oscillation and Mass***

Angle = $q$ = Length = L =

Table 2 Analysis of period of oscillation with changing mass

| Mass [ ] | Time for 5 Oscillations [ ] | Period $T_{exp}$ [ ] | Period $T_{cal-2}$ [ ] | PD $T_{exp}$ & $T_{cal-2}$ [ ] |
|---|---|---|---|---|
| | | | | |
| | | | | |
| | | | | |
| | | | | |

### *Part C: Investigation of Length effect and calculate gravitational constant*

Angle = $q$ = Mass = M =

Table 3 Analysis of period of oscillation with changing length

| Length [ ] | Trial 1 | | Trial 2 | | Trial 3 | | Trial 4 | |
|---|---|---|---|---|---|---|---|---|
| | 5 Osci. [ ] | $T_1$ [ ] | 5 Osci. [ ] | $T_2$ [ ] | 5 Osci [ ] | $T_3$ [ ] | 5 Osci. [ ] | $T_4$ [ ] |
| | | | | | | | | |
| | | | | | | | | |
| | | | | | | | | |
| | | | | | | | | |
| | | | | | | | | |
| | | | | | | | | |
| | | | | | | | | |
| | | | | | | | | |

Table 4 Investigation of gravitational acceleration

| Length L [ ] | Average Period $\overline{T}_{exp}$ [ ] | Mean deviation $\delta \underline{T}$ [ ] | Gravitational constant, $g_{exp}$ [ ] | PD 9.81 $m/s^2$ & $g_{exp}$ [ ] | $\overline{T}^2_{exp}$ [ ] |
|---|---|---|---|---|---|
| | | | | | |
| | | | | | |
| | | | | | |
| | | | | | |
| | | | | | |
| | | | | | |
| | | | | | |
| | | | | | |

- Find PD for followings:
  a) PD for $g_{avg_exp}$ vs $g$ = 9.81 $m/s^2$
  b) PD for $g_{avg_exp}$ vs $g_{slope}$
  c) PD for $g_{slope}$ vs $g$ = 9.81 $m/s^2$

# REFERENCES

1) Fundamentals of Physics by David Halliday, Robert Resnick and Jearl Walker, John Wiley Publication, 2018
2) University Physics, vol-1 by William Moebs, Samuel J. Ling, Jeff Sanny, OpenStax Publication, 2016 https://openstax.org/details/books/university-physics-volume-1
3) University Physics, by Harris Benson, John Wiley and Sons, Inc. 1996
4) Physics for Scientist and Engineers with Modern Physics, by Raymond A. Serway, Saunders College Publishing, 2004
5) University Physics, by Hugh D. Young, Addison-Wesley Pub. Co. 2004
6) Physics for Scientist and Engineers, Extended Version, by Fishbane, Gasiorowicz and Thornton, Prentice Hall, Inc. 2005
7) Physics for Scientist and Engineers with Modern Physics, Douglas A. Giancoli, Prentice Hall Publication, 2008
8) Principles and Practice of Physics, 1st edition by Mazur, Pearson Publication, 2006
9) Conceptual Physics, 12th edition by Paul G. Hewitt, Pearson Publication, 2015
10) oLabs Simulations, Amrita Vishwa Vidyapeetham and CDAC Mumbai, Ministry of Electronics and Information Technology, India, 2020, https://amrita.olabs.edu.in/
11) Open Educational Resource and Open Source Physics, Singapore, 2020, https://iwant2study.org/ospsg/index.php/interactive-resources/physics
12) Tracker Software, Douglas Brown, Robert Hanson and Wolfgang Christian, Open Source Physics, 2020, https://physlets.org/tracker/
13) PhET Interactive Simulations, University of Colorado, 2020, https://phet.colorado.edu/en/simulations/
14) oPhysics-Interactive Physics Simulations, Tom Walsh, 2020, https://ophysics.com/ https://ophysics.com/f3.html
15) Newton Law Simulator, Walter Fendt, 2018 https://www.walter-fendt.de/html5/phen/newtonlaw2_en.htm
16) The Physics Aviary, Fmcculley, 2020, https://www.thephysicsaviary.com/
17) GigaPhysics, Donovan Harshbarger, 2016, https://www.gigaphysics.com/labs.html https://www.gigaphysics.com/momentum_lab.html